\pdfoutput=1

\documentclass[%
 reprint,
 amsmath,amssymb,
 aps,
prb,
nofootinbib,
]{revtex4-2}

\usepackage{tabularx}
\usepackage{hyperref} 
\hypersetup{colorlinks=true, citecolor=blue, urlcolor=blue, linkcolor=blue} \usepackage{siunitx}
\usepackage{graphicx}
\usepackage{bm}

\begin{document}


\title{Prediction of stable Li-Sn compounds: \\ boosting \textit{ab initio}
searches with neural network potentials}

\author{Saba Kharabadze, Aidan Thorn, Ekaterina A. Koulakova, and Aleksey N.
Kolmogorov} 
\affiliation{
  Department of Physics, Applied Physics and Astronomy,\\ Binghamton University,         State University of New York, \\PO Box 6000, Binghamton, New York 13902-6000, USA.
}%

\begin{abstract} The Li-Sn binary system has been the focus of extensive
  research because it features Li-rich alloys with potential
  applications as battery anodes. Our present
  re-examination of the binary system with a combination of machine
  learning and {\it ab initio} methods has allowed us to screen a
  vast configuration space and uncover a number of
  overlooked thermodynamically stable alloys. At ambient pressure, our
  evolutionary searches identified a new stable Li$_3$Sn phase with a
  large BCC-based hR48 structure and a possible high-$T$ LiSn$_4$
  ground state. By building a simple model for the observed and
  predicted Li-Sn BCC alloys we constructed an even larger viable hR75
  structure at an exotic 19:6 stoichiometry. At 20~GPa, new 11:2, 5:1,
  and 9:2 phases found with our global searches destabilize previously
  proposed phases with high Li content. The findings showcase the
  appreciable promise machine learning interatomic potentials hold for
  accelerating {\it ab initio} prediction of complex materials.

\end{abstract}

\maketitle


\section{Introduction}

Accurate density functional theory (DFT) approximations introduced over the
past three decades have proven to be powerful practical solutions for quantum
mechanical description of materials properties
\cite{hautier2012computertolab,oganov2019,2019roadmap}. Reliable prediction
of structures' thermodynamic stability \cite{ak28,ak09,sun2016metastable} has
had a particularly transformative impact on the materials discovery process, as
{\it ab initio} ground state searches have been used to screen large chemical
spaces and identify new synthesizable materials \cite{materialsproject,ak28}.
Despite the growing number of confirmed predictions
\cite{hautier2012oxides,Hajinazar2021,lepeshkin2018}, the pace of 
DFT-guided discovery is ultimately limited by the high cost of {\it ab initio}
calculations.

Machine learning potentials (MLPs) have emerged as an attractive alternative approach for describing
interatomic interactions \cite{behler2016perspective,bartok2017}. Being flexible classical models, they
can fit DFT potential energy surfaces (PESs) to within the typical systematic errors of the reference
method but operate at a fraction of the DFT computational cost
\cite{Hajinazar2021,zuo2020performance,behler2017,deringer2019}. These capabilities have expanded
the scope of atomistic simulations well beyond what has been possible with {\it ab initio} methods. For
example, extended MLP-based molecular dynamics have been employed for systems with 10$^5$ atoms
\cite{ deringer2021}. MLP simulations have been used to examine phase transitions \cite{Behler2008,
Khaliullin2010,Eshet2012,Sosso2012, Baldock2016, Mocanu2018, Kruglov2019}, probe complex surface/interface 
geometries \cite{ Artrith2013, bartok2017, Boes2017}, simulate crack propagation \cite{shen2021},
analyze large-scale defects \cite{stricker2020machine, goryaeva2021}, model disordered materials
\cite{deringer2021, qian2019thermal}, and study other complex systems overviewed by
\citet{Hart2021}.

MLP-assisted prediction of ground state structures differs notably
from these materials modeling applications. While ns-long MD
trajectories or a-few-nm-sized configurations cannot be checked with
DFT directly, pools of candidate structures with lowest Gibbs free
energy found in MLP-based searches can and should be examined with the
reference method. The benefit of using surrogate models should then be
measured against the net cost of their parametrization and the
following DFT analysis of viable candidates. MLPs have proven to be
essential for finding stable nanoparticle configurations because
direct global optimizations with the DFT become impractical for
cluster sizes above a few dozen atoms \cite{Heiles2013, Jager2018,
  Baletto2019}, while empirical potentials generally lack the accuracy 
needed to effectively guide {\it ab initio}
searches. In fact, our recent large-scale comparative study showed a
significant advantage of a typical neural network (NN) interatomic
model over traditional potentials and led to the revision of several
{\it ab initio} ground states for Au clusters with 30-80 atoms
\cite{ak40}.

The advantage of developing and using MLPs in the
exploration of crystalline solids is less apparent because ordered phases tend
to have relatively small unit cells. To the best of our knowledge, MLP-guided
global structure searches for crystalline ground states have been performed
only in a handful of studies \cite{Hart2019, ak37, Deringer2018, Huang2018, Podryabinkin2019, Yang2021, Deringer2017, Deringer2017b, Podryabinkin2019, Behler2008, Csanyi2018, Deringer2018b, Deringer2020} dedicated primarily to unary materials \cite{Deringer2018, Huang2018, Podryabinkin2019, Yang2021, Deringer2017, Deringer2017b, Podryabinkin2019, Behler2008, Csanyi2018, Deringer2018b, Deringer2020}. This work has resulted
in very few predictions of new thermodynamically stable phases, such as
Co-Nb-V, Al-Ni, and Al-Ni-Ti \cite{Hart2019} and several Mg-Ca
alloys \cite{ak37}, as investigations of elemental boron \cite{
Deringer2018, Huang2018, Podryabinkin2019, Yang2021}, carbon \cite{
Deringer2017, Deringer2017b, Podryabinkin2019}, silicon \cite{Behler2008,Csanyi2018}, and
phosphorous \cite{Deringer2018b, Deringer2020} have either reproduced previously
known structures or found metastable structures. 

In this study, we aim to illustrate the capabilities of our recently
developed framework for MLP-acccelerated structure prediction
\cite{ak34,Hajinazar2021} by performing a large-scale exploration of
Li-Sn alloys. The motivation for (re)investigating this common binary
system is twofold. First, the existence of several Li-rich compounds
with high electrical conductivity and better ductility compared to
other group-XIV elements make them appealing candidate materials for
Li-ion battery anodes \cite{Sen2017,Li2013,zhang2015}. Second, the
binary system has been recently investigated with the state-of-the-art
{\it ab initio} prediction methods in two systematic studies and shown
to host new thermodynamically stable compounds at ambient and elevated
pressures \cite{Sen2017,Mayo2017}. The development and application of
an accurate NN model for Li-Sn has allowed us to probe over a million
structures across the full composition range at different ($P$,$T$)
conditions and identify several new ground states. Particularly
surprising are thermodynamically stable BCC-based Li$_3$Sn and
Li$_{19}$Sn$_6$ phases with large prototypes that appear not to have
been observed in any other binary alloys. These findings illustrate
the appeal of employing MLPs to accelerate {\it ab initio} prediction
of complex materials.

\section{Methodology}

\subsection*{Density functional theory calculations} All DFT calculations were
performed with the Vienna \textit{ab-initio} simulation package ({\small
VASP}) \cite{VASP,VASP2,VASP3,VASP4}. To ensure proper description of materials
under high pressures, we chose projector augmented wave potentials \cite{PAW}
with semi-core electrons of Li (1$s$) and Sn (4$d$). Unless specified otherwise, we
used the Perdue-Burke-Ernzerhof (PBE) exchange-correlation functional \cite{PBE}
within the generalized gradient approximation (GGA) \cite{Langreth1983} and the
energy cutoff of 500 eV. Select phases were examined with a higher 700 eV
energy cutoff and within the local density approximation (LDA) \cite{LDA1,LDA2}
or strongly-constrained and appropriately-normed (SCAN) in the meta-GGA
\cite{SCAN}. All crystalline structures were evaluated with dense ($\Delta k
\sim 0.02$ $\si{\angstrom}^{-1}$) Monkhorst-Pack $k$-point meshes
\cite{Monkhorst1976}.

\subsection*{Neural network interatomic potential} A NN model of the
Behler-Parrinello type was constructed using an automated iterative scheme
implemented in our {\small MAISE-NET} framework \cite{Hajinazar2021} that
features an evolutionary sampling algorithm to generate representative
reference structures and a stratified training protocol to build multicomponent
models on top of elemental ones \cite{ak34}. We relied on our previously
developed Li and Sn NN models \cite{Hajinazar2021} with 51-10-10-1
architectures and $\sigma^{\text{Li}}_{E}=2.3$~meV/atom and
$\sigma^{\text{Sn}}_{E}=8.3$~meV/atom root-mean-square errors (RMSEs). The
Li-Sn NN with a 145-10-10-1 architecture and 1,880 adjustable parameters 
\footnote{Each of the two elemental NNs was expanded with 940 adjustable 
interspecies weights connecting the 8 pair and 43 triplet input symmetry 
functions with 10 neurons in the first hidden layer:
$(8_{\text{LiSn}}+43_{\text{LiLiSn}}+43_{\text{LiSnSn}}) \times 10_{\text{Li}} + (8_{\text{SnLi}}+43_{\text{SnSnLi}}+43_{\text{SnLiLi}}) \times 10_{\text{Sn}}$.} 
was fitted to 6,046 energy and 46,410 force data in binary structures with up to 32
atoms generated in four {\small MAISE-NET} cycles. The optimization of the binary
model for 120,000 steps resulted in $\sigma^{\text{Li-Sn}}_{E}=10.2$~meV/atom
and $\sigma^{\text{Li-Sn}}_{F}=49.0$~meV/\si{\angstrom} RMSEs over a test set
of 671 structures. All elemental and binary models were based on our stardard
set of Behler-Parrinello symmetry functions with a 7.5 \si{\angstrom} cutoff
radius \cite{ak40}. In total, the extension of the elemental Li and Sn models 
to the binary one cost under 35,000 CPU hours.

\begin{figure*}[!ht] \centering
  \includegraphics[width=\textwidth]{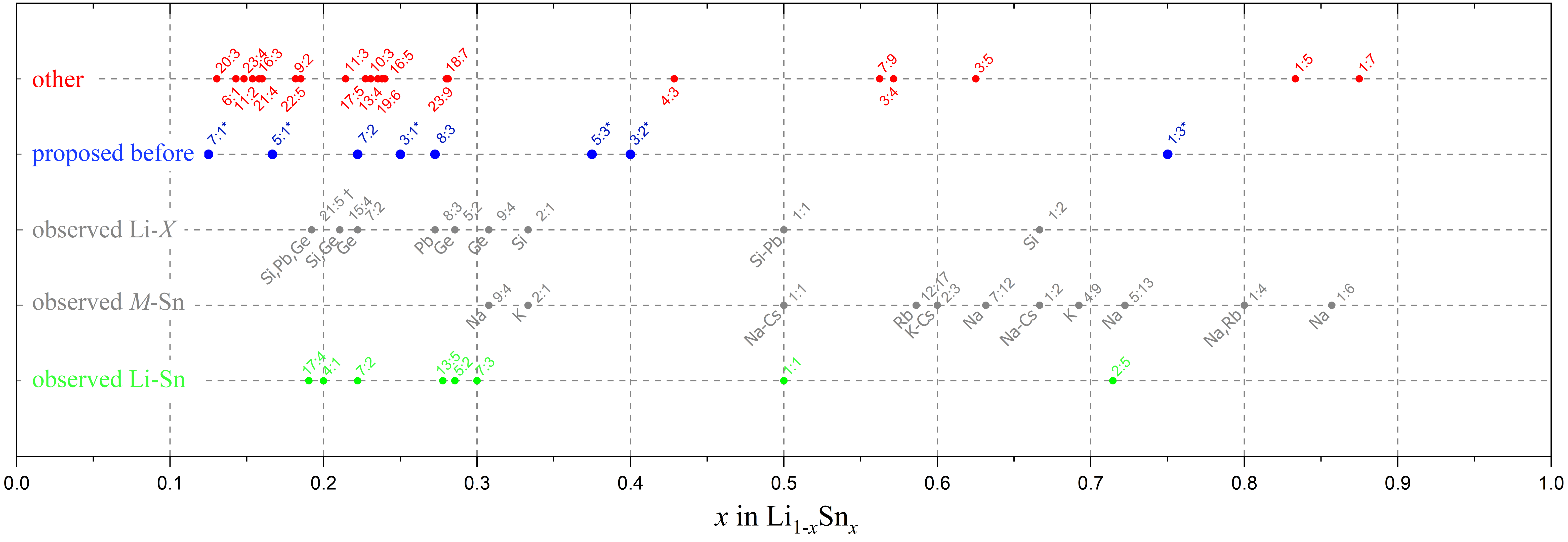} \caption{ Relevant
  compositions in the Li-Sn system explored with our NN-based evolutionary
  searches. The bottom green set denotes experimentally observed Li-Sn
  compounds \cite{ICSD}. The two grey groups specify stoichiometries
  observed in related \textit{M}-Sn and Li-\textit{X} binaries \cite{ICSD}.
  The blue set refers to thermodynamically stable compounds predicted in
  previous studies \cite{Sen2017,Mayo2017}  for 0~GPa and 20~GPa (the latter
  are marked with asterisks). The top red group corresponds to additional
  compositions considered in this study: all possible $m:n$ ratios with $(m+n)
  \leq 8$ and a few larger subsets with $m$ = 6-23 and $n$ = 1-9 near regions
  with known stable compounds.  } 
\label{fig:phases} 
\end{figure*}

\subsection*{Structure prediction strategy} {\it Overview} Identification of
{\it ab initio} Li-Sn ground states at different $(P,T)$ conditions included
evolutionary searches with the NN model at $T=0$~K, re-examination of viable
$T=0$~K candidates with the DFT, and investigation of phase stability at
elevated temperatures at the NN and DFT levels. The global search for stable
Li-Sn alloys at $T=0$~K and $P=0$~GPa or $P=20$~GPa was driven by an
evolutionary algorithm implemented in {\small MAISE} \cite{Hajinazar2021}.
Detailed in our previous studies \cite{ak16,ak23,Hajinazar2021}, this global
optimization engine has been used primarily in combination with DFT to
predict or solve complex ground states. It has led to the discovery of several
confirmed new prototypes with large unit cells: oP10-FeB$_4$ \cite{ak16,ak26},
mP20-MnB$_4$ \cite{ak28}, and tI56-CaB$_6$ \cite{ak23} found without any
structural input as well as mP24-Cu$_2$IrO$_3$ and
tP56-Na$_3$Ir$_3$O$_8$ found by seeding the searches with known
structures \cite{ak42,ak36}. Noteworthy features of the present NN-based
evolutionary searches and post-search analysis are described below.

{\it Selection of compositions} Exhaustive screening of relevant
stoichiometries is a requisite for identification of stable compounds. It has
been demonstrated \cite{zunger2009,oganov2014} that unsupervised
variable-composition evolutionary searches can result in an efficient location
of stable stoichiometries. One of the recent studies on Li-Sn \cite{Sen2017}
employed such an algorithm and refined solutions with follow-up evolutionary
searches at select compositions. Given the richness of the Li-Sn phase diagram
and the complexity of the known ground states at non-trivial compositions, we
relied on a supervised selection of stoichiometries for our fixed-composition
searches.

Figure \ref{fig:phases} summarizes available information on Li-Sn and related
chemical systems helpful for determining compositions pertinent to this study.
The three lowest sets list previously synthesized Li-Sn, $M$-Sn ($M$ = Na-Cs),
and Li-$X$ ($X$ = Si, Ge, and Pb) alloys. Among the $M$-Sn binaries, Li-Sn
stands out as a system with multiple Li-rich compounds and unique stable
compositions across the full range. It is evident that the size of the alkali
metals is a dominant factor defining stable $M$:Sn ratios, which renders the
observed Na-Cs tin alloy compositions not particularly relevant for Li-Sn under
ambient conditions. The Li-$X$ set, on the other hand, provides important clues
regarding possible compositions and morphologies that could be observed in the
Li-Sn binary. As discussed in the following sections, some of the previously
predicted compounds displayed near the top of Fig. \ref{fig:phases} were shown to
be stable in these prototypes.

Our main focus was on compositions with previously reported analogs in Fig.
\ref{fig:phases} containing fewer than 40 atoms in the primitive unit cells.
The very large observed cF420-Li$_{17}$Sn$_4$ prototype was examined directly
with the DFT, as finding global minima from scratch for crystalline structures
above about 40 atoms is challenging even with evolutionary searches based on
surrogate models. We also considered all possible $m:n$ ratios with $m+n\leq 8$
and included a few larger $m:n$ subsets with $m$ = 6-23 and $n$ = 1-9 to sample the
Li-rich end of the phase diagram (the top row in Fig. \ref{fig:phases}).

{\it Evolutionary search settings} At each selected composition, we ran
separate evolutionary runs with different numbers of formula units. Typically,
structures had between 1 and 8 formula units and did not exceed 40 atoms in the
primitive unit cell. Randomly generated populations of 32 members were evolved
for up to 300 generations with standard evolutionary operations. Namely, 8 new
members were constructed from a single parent structure through mutation
(random atom displacements, atom swaps, and unit cell distortions) while 24
offspring were created from two parents through crossover (combination of two
roughly equal parts obtained with planar cuts) \cite{ak16}. Child structures
were locally relaxed with the NN potentials for up to 300
Broyden–Fletcher–Goldfarb–Shanno (BFGS) minimization steps and assigned a
fitness based on the final enthalpy. Our fingerprint method based on the radial
distribution function (RDF) \cite{ak23, Hajinazar2021} was used to identify
similar structures and decrease their survival probability.

{\it Post-search DFT analysis at $T=0$~K} Upon completion of a NN-based
evolutionary run, a subset of viable candidates from all generated local minima
was selected for further re-examination with the DFT.  The global DFT minimum
is captured provided that it is (i) at least a local minimum on the NN PES;
(ii) visited by the search algorithm; and (iii) included in the analysis pool
\cite{ak40}. The main adjustable parameter in the selection process is the
enthalpy window. By default, we collected all distinct minima within
20~meV/atom ($\approx 2\sigma^{\text{Li-Sn}}_{E}$) above the NN ground state to
account for the NN typical errors in the evaluation of relative enthalpies but
reduced the window at some compositions with a large number of competing
low-symmetry states. We relied on a 0.95 RDF-based dot product cutoff to
exclude similar structures and used a 0.1 tolerance to symmetrize the unit
cells \cite{Hajinazar2021} before relaxing them with DFT. 

{\it Analysis of phase stability at high $T$} Identification of
high-$T$ ground states is commonly done by including the vibrational
entropy contribution to the Gibbs free energy for a pool of
low-enthalpy phases found in global structure searches at $T=0$~K
\cite{Hajinazar2021,oganov2019}. The relative change in Gibbs free
energy can reach a few dozen meV/atom for structures with
substantially different phonon density of states (DOS) observed,
\textit{e.g.}, in $\alpha$-Sn and $\beta$-Sn
\cite{pavone1998,legrain2016,mehl2021}. Given the demanding nature of
phonon calculations at the DFT level, we used the following protocol
to examine high-$T$ phase stability. Firstly, we calculated thermal
corrections with the NN model ($\Delta F_{\text{vib}}^{\text{NN}}$)
for all distinct candidate structures in the $T=0$~K DFT pools within
a 20 meV/atom window. Secondly, we constructed a new convex hull at
600~K using $G^{\text{DFT+NN}}=H^{\text{DFT}}+\Delta
F_{\text{vib}}^{\text{NN}}$ and selected up to 5 structures per pool
with Gibbs free energies no further than 10 meV/atom from the
tie-line. Finally, we evaluated thermal corrections for these
structures with DFT and determined high-$T$ ground states by examining
convex hulls at all temperatures in the 0-800~K range, in intervals of
10~K.  Both NN and DFT phonon calculations were performed with the
finite displacement method as implemented in Phonopy
\cite{phonopy}. We used symmetry-preserving expansions of primitive or
conventional unit cells to generate supercells with 72-216
atoms. Structures included in the thermodynamic stability analysis
were dynamically stable, and any structures which remained dynamically
unstable were disregarded if a new structure could not be created from
their imaginary frequency eigenmodes.

\section{Results and Discussion}

\begin{figure}[t] \centering
  \includegraphics[width=0.48\textwidth]{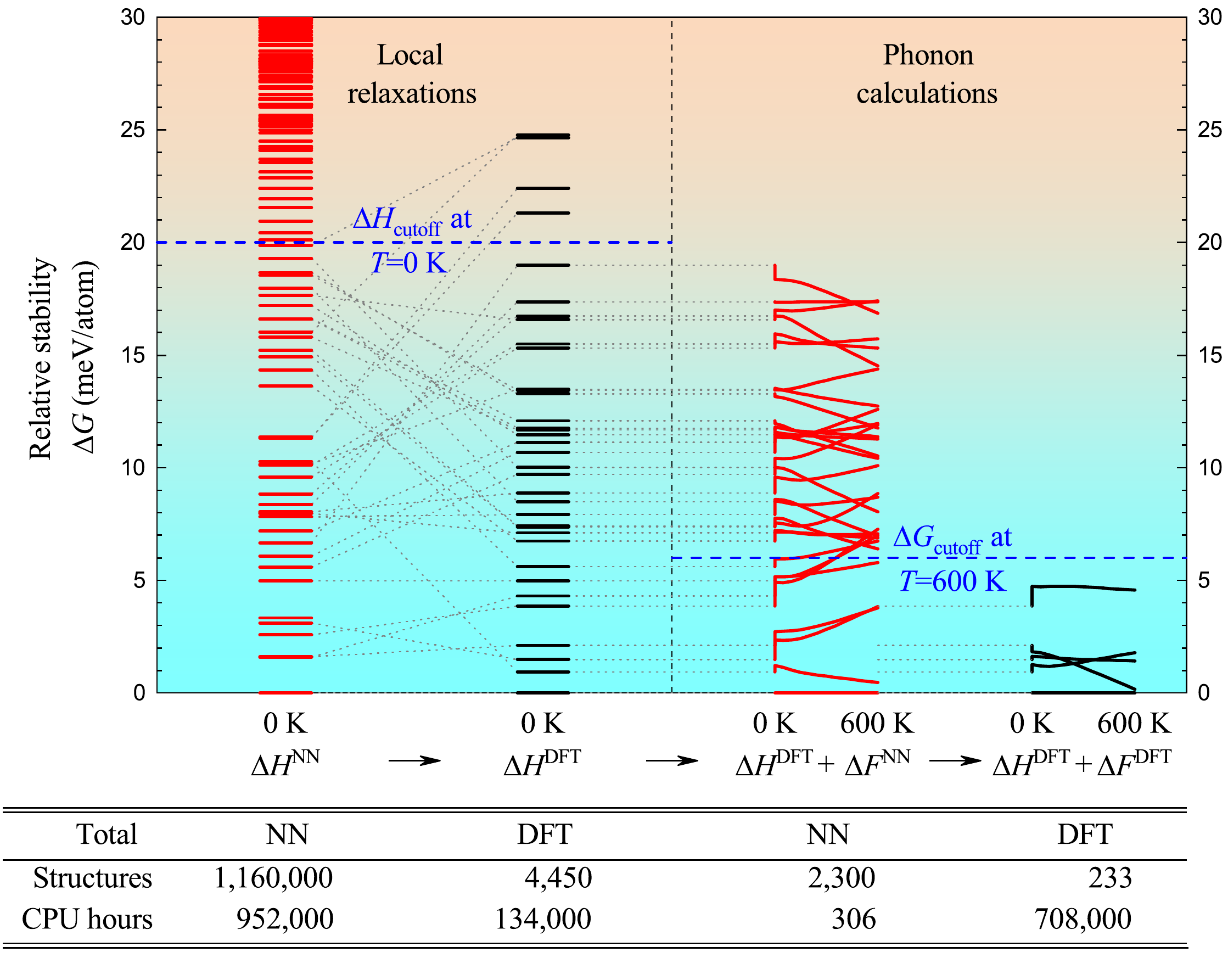} \caption{ An
  overview of the NN-guided identification of zero-$T$ and high-$T$ \textit{ab
  initio}  ground states. The flowchart illustrates the selection of viable
  candidates following a representative evolutionary search performed for
  Li$_{18}$Sn$_4$ unit cells at 0 GPa. The two sets on the left correspond to
  the lowest-enthalpy NN minima (in red) and the resulting reoptimized DFT
  minima (in black). The two subsets on the right show the Gibbs free energies
  in the 0-600 K range with the phonon entropy contributions calculated with
  the NN model (in red) and the DFT (in black). The table at the bottom shows
  the tally of the structures examined and computational resources used in the
  full exploration of the Li-Sn binary.   } 
\label{fig:flow} 
\end{figure}

  \subsection{Overview of global structure search results}

\begin{figure}[t] \centering
  \includegraphics[width=0.48\textwidth]{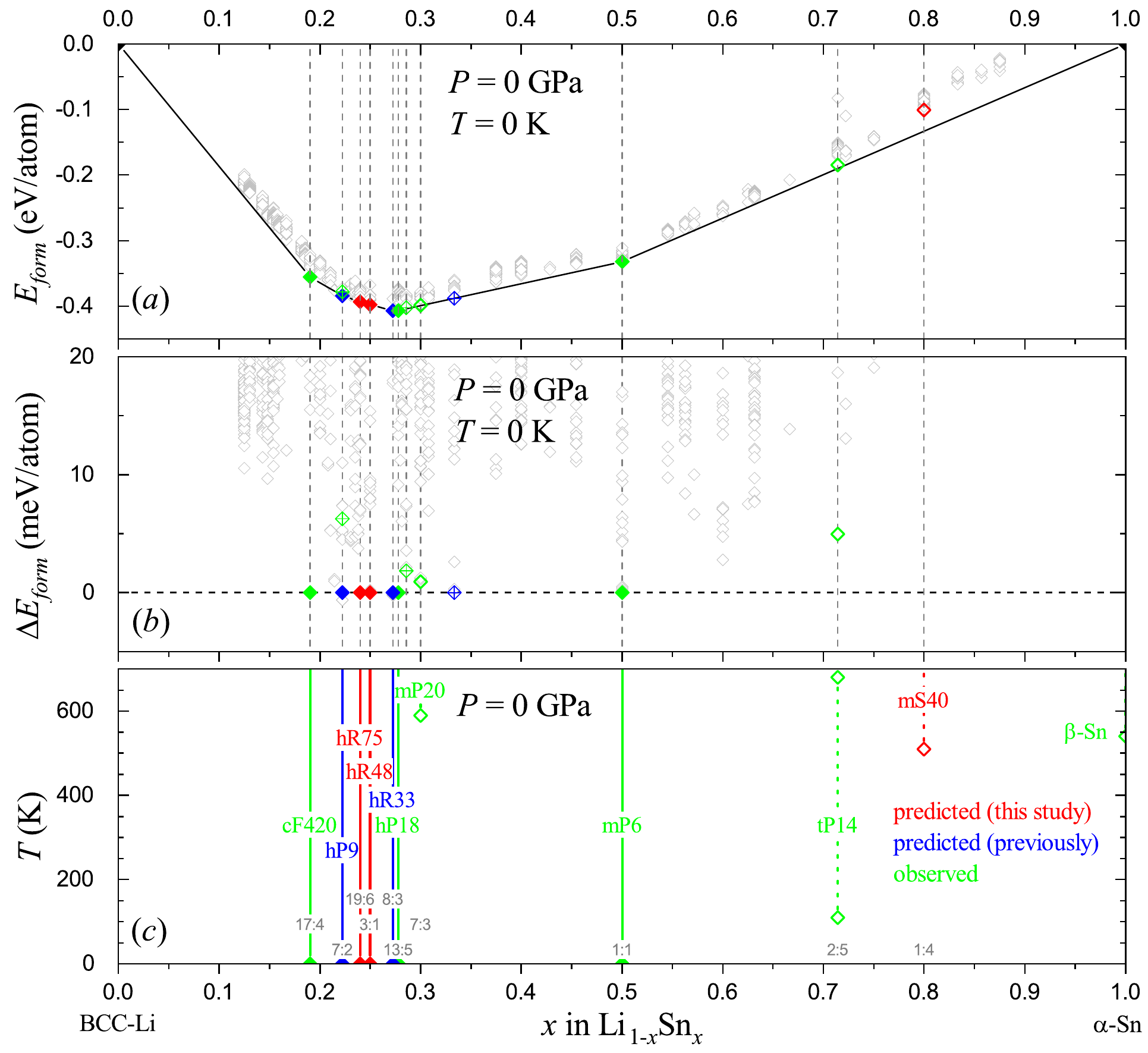} 
  \caption{ DFT thermodynamic stability analysis of Li-Sn alloys at $P=0$~GPa.
  The Li-B phases observed experimentally, proposed previously \cite{Mayo2017,
  Sen2017}, and predicted in this study are shown in green, blue, and red,
  respectively. The solid (open) symbols denote phases stable at zero
  (elevated) temperatures, while the crossed symbols correspond to phases that
  remain metastable at all considered temperatures. (a) The convex hull at
  $T=0$~K.  (b) The distance from the convex hull at $T=0$~K. (c) The
  temperature range of phase stability. The new thermodynamically stable phases
  shown in red are hR75-Li$_{19}$Sn$_6$, hR48-Li$_3$Sn, and mS40-LiSn$_4$.  
  }
\label{fig:hull00} 
\end{figure}

\begin{figure}[t] \centering
  \includegraphics[width=0.48\textwidth]{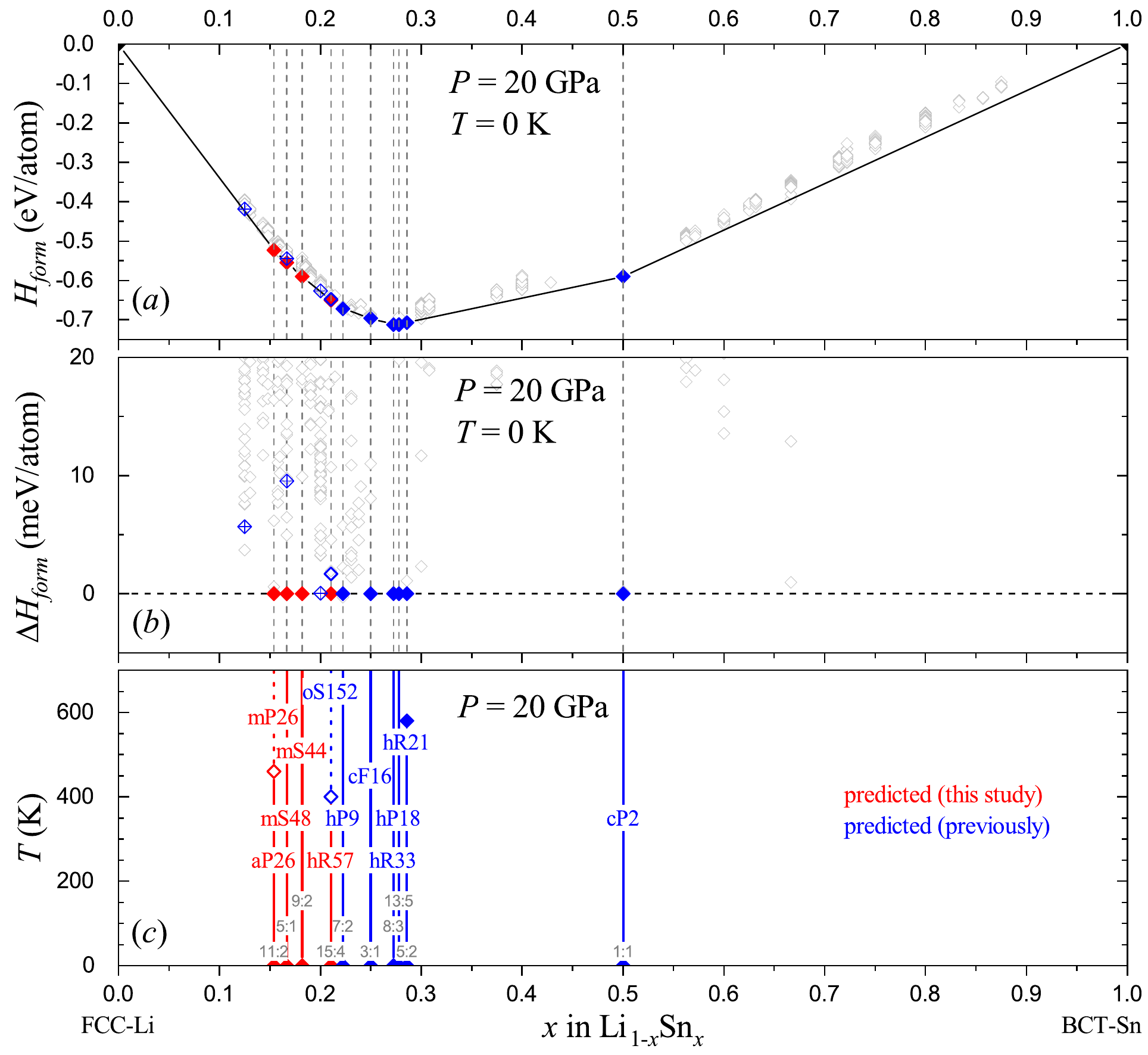} \caption{ The
  same as in Fig. \ref{fig:hull00}, but at $P=20$~GPa. The four
  putative ground states in red identified with our NN-based evolutionary
  searches redefine the Li-rich end of the previously proposed $T=0$~K convex
hull.  } 
\label{fig:hull20} 
\end{figure}

The information presented in Figs.~\ref{fig:flow}-\ref{fig:hull20}, serves to
illustrate the size of the configuration
space sampled in this work and the reliability of the NN-based structure
prediction approach employed for identification of stable Li-Sn phases.

According to our results, populations in the evolutionary searches typically
stopped improving after 210, 910, and 2,300 configurations were probed in runs
with up to 14 atoms, 15-24 atoms, and over 25 atoms per unit cell,
respectively. While the stochastic searches are not guaranteed to converge to
the global minimum in every instance, we find that sampling 4,400 candidates
structures with up to 24 atoms is generally sufficient. Overall, more than 1.1
million local optimizations with the NN model were performed for binary
structures with no initial symmetry and an average of 19.4 atoms per unit cell,
which required 952,000 CPU hours (see Fig.~\ref{fig:flow}). An equivalent set
of DFT calculations would require an estimated 100 million CPU hours (the
increase in the average computational cost was estimated by relaxing 20
structures with 5-30 atoms for the same number of steps in the NN and DFT
optimizations). The selection and symmetrization of configurations with low NN
enthalpies narrowed the pool down to 4,450 structures (about 13
structures per pool) and cost considerably less to re-optimize with
DFT. Importantly, the near-stable structures were described more accurately
with the NN model (6.9~meV/atom RMSE compared to ${\sigma^{\text{Li-Sn}}_{E}=10.2}$
~meV/atom) and underwent relatively small geometrical changes in the
following DFT local relaxations (2.0~meV/atom enthalpy decrease on average).
As a result, the best DFT minima ranked among the most favored ones within the
NN pools (the structures with the lowest NN enthalpy remained favored at the
DFT level in 45\% of all considered cases) and had either matching or lower
enthalpies compared to all previously reported Li-Sn phases at ${T=0}$~K.

Our analysis of phase stability at elevated temperatures benefited similarly
from the use of the NN model. Following the selection scheme outlined in
Section II, we performed phonon calculations for 2,300 and 233 structures at
the NN and DFT levels, respectively. With the free energy contributions
averaging 157~meV/atom at 600~K, the correspondence between the two
methods was 7.2~meV/atom. With the RMSE of 1.5~meV/atom and the 4.9~meV/atom
average change in {\it relative} stability at $T=600$~K (see Fig. S1 for
typical results), the NN-based screening allowed us to examine an order of
magnitude more candidate structures and uncover possible high-$T$ ground states
at 0~GPa and 20~GPa.

It is important to note that while DFT systematic errors in calculations of
thermodynamic stability may reach a few dozen meV/atom in certain cases
\cite{ak16,ceder2006oxidation}, relative energies evaluated for similar
structures tend to be far more accurate due to cancellation of errors
\cite{ak06,ak37}. Our tests performed for different functionals and
convergence settings (Tables S1 and S2) demonstrate that the relative enthalpies
can be resolved to within a fraction of 1~meV/atom for most considered
structures with underlying BCC morphologies. Temperature estimates
for transitions between near-degenerate phases with small differences in free
energy corrections can have significant uncertainties of a few hundred Kelvin
\cite{ak31,ak37}.

Our main findings regarding competing Li-Sn phases at 0~GPa and 20~GPa are
summarized in Figs. \ref{fig:hull00} and \ref{fig:hull20}. The DFT results
  allowed us to refine known ${T=0}$~K convex hulls (top), identify additional
metastable alloys (middle), and establish temperature ranges of phase stability
(bottom). Namely, two new ambient-pressure ground states are added next to the
previously proposed ones \cite{Sen2017,Mayo2017} and several new high-pressure
ground states substantially redefine the Li-rich end of the convex hull
predicted in Ref. \onlinecite{Sen2017}. The collection of local minima obtained in
our NN-based evolutionary searches illustrates the presence of many more phases
within the 20~meV/atom window than found with the DFT-based random sampling
approach \cite{Mayo2017}. The inclusion of thermal corrections indicates that
only a few considered metastable phases have a chance of becoming
thermodynamically stable.

\subsection{Stability and morphology of Li-Sn phases}

The variety of observed Sn alloy morphologies \cite{jing2016, cheng2017,
salamat2013, wang2011} can be attributed to the particular position of the
element in the periodic table. Similarly to the C-Ge group-XIV members, Sn
displays a propensity to forming covalent bonds and crystallizes in the open
diamond structure ($\alpha$-Sn) at low $(P,T)$. In contrast to the light
isoelectronic elements, the more pronounced $s$-$p$ relativistic splitting
reduces the favorability of the $s$-$p$ hybrid orbitals and, consequently, Sn
adopts various ground states with metallic bonding, such as $\beta$-Sn above
286~K \cite{pavone1998} or closed BCT structure between 11~GPa and 32~GPa
\cite{jing2016,salamat2013}. Li is an archetypical $sp$ metal with the BCC
ground state under ambient conditions transforming into FCC above $\sim$~5~GPa
\cite{guillaume2011, hutcheon2019}.

\begin{figure*}[t] \centering
  \includegraphics[width=0.96\textwidth]{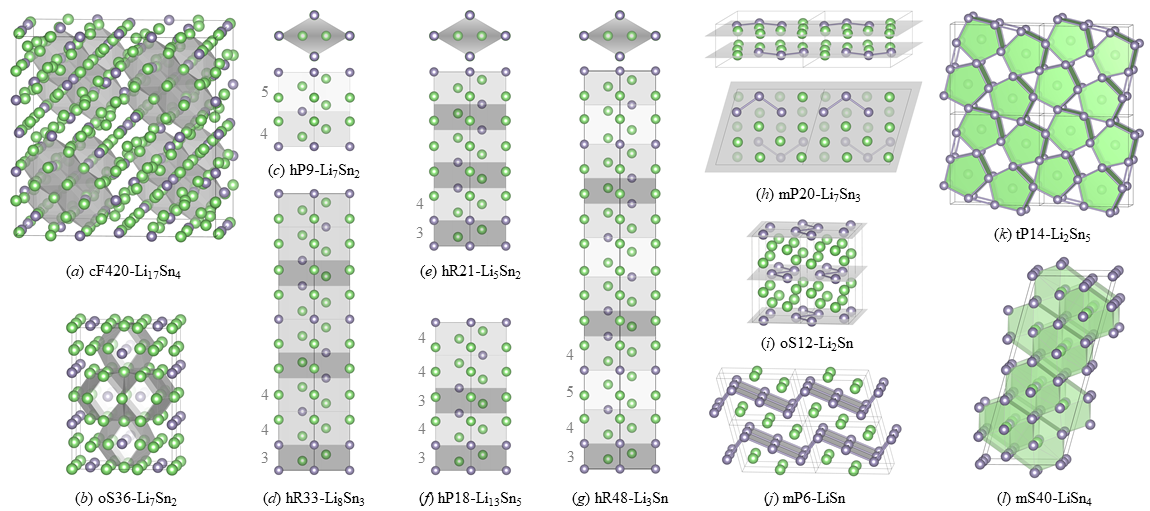} 
  \caption{
    Ambient-pressure Li-Sn phases observed experimentally (a,b,e,f,h,i,k),
    proposed previously in Refs. \onlinecite{Mayo2017,Sen2017} (c,d), and
    predicted  in this study (g,l). The Li and Sn atoms are shown in green and
    grey, respectively, while the shaded polyhedra, sections, and planes
    highlight distinctive morphological features discussed in the text. The
    [111] BCC alloys shown in (c-g), along with our predicted
    hR75-Li$_{19}$Sn$_6$ not included due to its large size, can be uniquely
    defined with sequences of Sn blocks, \textit{e.g.}, $\vert$3454$\vert$ for
    hR48-Li$_3$Sn. These images were generated with the VESTA package \cite{VESTA}.
  }
\label{fig:LiSnStrs} 
\end{figure*}

In combination, Li and Sn have been observed to form BCC-based alloys at high
Li concentrations and compounds with covalent frameworks at high Sn
concentrations. The prevalence of BCC motifs can be viewed as a consequence of
the negative chemical pressure induced by the larger Sn ($a^\text{Sn}_\text{BCC}=3.81 
\si{\angstrom}$) that further favors the BCC-Li lattice ($a^\text{Li}_\text{BCC}=3.43 
\si{\angstrom}$) over the high-pressure FCC-Li polymorph.
The largest magnitude of the formation enthalpy achieved around the 4:1
composition (Figs. \ref{fig:hull00} and \ref{fig:hull20}) is consistent with the 
octet rule \cite{Sen2017,octet1, octet2, octet3}.

At ambient pressure, Li$_{17}$Sn$_4$ is the most Li-rich synthesized alloy in
this binary system. Goward \textit{et al}. performed a comprehensive XRD analysis of
Li$_{22}X_5$ ($X$ = Ge, Sn, and Pb) compounds and determined that the slight
compositional variability across the series originates from selected occupation
of Li sites \cite{goward2001}. The ordered cF420 ($F\bar{4}3m$) structural model used to
simulate Li$_{17}$Sn$_4$ \cite{Sen2017, Mayo2017} has only one Sn Wyckoff site
($16e$) with the full (8+6)-atom BCC coordination that produces tetrahedral
agglomerates of Sn polyhedra (see Fig. \ref{fig:LiSnStrs} (a)). The other 
three Sn Wyckoff sites, $16e$, $24g$, and $24f$, are surrounded by 13 Li atoms 
and generate isolated Sn atoms sufficient to fill up all space with interlocking 
polyhedra. As in previous studies \cite{Mayo2017,Sen2017}, the phase is found 
to be thermodynamically stable in our DFT calculations.

The observed oS36-Li$_7$Sn$_2$ phase represents a regular BCC lattice with
fully occupied sites. The Sn atoms in this Li$_7$Ge$_2$ prototype are not fully
segregated, making one of the two Sn Wyckoff sites have a Sn neighbor in the
second shell at 3.20 \AA\ (Fig. \ref{fig:LiSnStrs} (b)). Interestingly, previous {\it ab initio}
studies revealed a lower-energy decoration of the BCC lattice at this
composition \cite{Mayo2017,Sen2017}. The simpler hexagonal hP9 phase with the
Li$_7$Pb$_2$ prototype does not contain directly interacting Sn atoms and ends
up being $\sim 6$~meV/atom more stable in (semi)local DFT approximations
\cite{Mayo2017,Sen2017, genser2001} and Table S1). Even though the energy
difference reduces to $\sim 3$~meV/atom at the SCAN level (Table
S1), the oS36 phase appears to be only metastable under typical synthesis
conditions. We reproduce the previously reported finding \cite{Sen2017} that
the inclusion of vibrational entropy changes the relative free energy
insignificantly (by less than 1~meV/atom up to 800~K in Fig. S2), which is not unexpected
given the similarity of the oS36 and hP9 morphologies.

The stable hR33-Li$_8$Sn$_3$ phase \cite{Mayo2017,Sen2017} with the
Li$_8$Pb$_3$ prototype has been proposed  \cite{Sen2017} to explain the nominal
compound synthesized in 1996 by Gasior \textit{et al}. \cite{gasior1996}. Just as
hP9-Li$_7$Sn$_2$, it exhibits an ordered pattern along the [111] 
crystallographic direction of the BCC lattice and can be represented with a
hexagonal unit cell (Fig. \ref{fig:LiSnStrs} (d)). In contrast to hP9-Li$_7$Sn$_2$, it has a
peculiar distribution of the Sn atoms: as discussed in Refs. \onlinecite{Mayo2017,
Sen2017}, the occupation of neighboring sites creates Sn-Sn dimers with an
unusually short 2.90 \AA\ interatomic distance.

The known hP18-Li$_{13}$Sn$_5$ and hR21-Li$_5$Sn$_2$ phases at nearby
compositions bear a strong resemblance to hR33 and contain Sn-Sn dimers of
length 2.91 \AA{} (on two of the three Sn Wyckoff sites) and 2.92 \AA\ (on the
only $6c$ Sn Wyckoff site), respectively. Our PBE calculations agree with the
previous results \cite{Sen2017} regarding the thermodynamic stability of the
former and the metastability of the latter. As summarized in Table S1,
hR21-Li$_5$Sn$_2$ is 1.8~meV/atom above the
{hP18-Li$_{13}$Sn$_5$ $\leftrightarrow$ mP6-LiSn} tie-line in this approximation but
marginally breaks the convex hull by $-0.4$~meV/atom in the SCAN calculations.
The higher sensitivity of the relative stability to the DFT flavor in this case
is likely caused by the fairly different morphology of the reference mP6-LiSn
phase discussed below.

In order to systematize the description of such BCC structures featuring the
same hexagonal base, we introduce a convenient notation that builds on the
detailed structural analysis presented in Ref. \onlinecite{Sen2017}. As can be
seen from Fig. \ref{fig:LiSnStrs}(c-g), Sn atoms 3, 4, and 5 layers apart along
the $c$ axis are shifted by 0, 1/3, and 2/3 unit cell fractions within the
$a$-$b$ base, respectively.  Therefore, Sn sequences can be used to uniquely
specify these trigonal crystal system structures and easily deduce whether they
belong to the hexagonal or rhombohedral lattice system.  Indeed, sequences with
the total number of Sn layer vertical separations divisible by three
(\textit{e.g.}, $\vert$45$\vert$ with 2 Sn atoms in the 9-atom primitive unit
cell) result in no net lateral shift and, hence, correspond to hexagonal
structures. All other sequences (\textit{e.g.}, $\vert$34$\vert$) must be
tripled to ensure orthogonality of the $c$ axis with the base and, thus, define
rhombohedral structures.

At the 3:1 composition, only metastable ambient-pressure phases have been
proposed so far \cite{Mayo2017,Sen2017} but our evolutionary searches for 12:4
unit cells uncovered a new thermodynamically stable member of the BCC family,
hR48-Li$_3$Sn with the $\vert3435\vert$ sequence. The rhombohedral structure
with a 16-atom unit cell appeared in both NN evolutionary runs
after 849, and 1,491 local optimizations, although it placed 17~meV/atom above
a cF16 BCC phase and ranked 4th in the NN pool. In the considered DFT
approximations, it was found to be the lowest-energy 3:1 configuration and
stable with respect to the decomposition into the previously proposed
hP9-Li$_7$Sn$_2$ and hR33-Li$_8$Sn$_3$ BCC-based ground states by about $-2$
~meV/atom at $T=0$~K (Table S1). Following phonon calculations indicated that
hR48-Li$_3$Sn should remain stable at elevated temperatures.

The finding motivated us to examine BCC structures more closely for other
possible ground states. The cluster expansion method has been widely used to
explore combinatorially large configuration spaces on given lattices
\cite{blum2004, kadkhodaei2021cluster} but an exhaustive screening of all possible
decorations of large unit cells would require a separate dedicated study.
Fortunately, the observed favorability of the [111] BCC motifs allowed us to
focus on the most promising subspace. Instead of relying on traditional atomic
clusters, we expanded the total energy as a linear combination of various
blocks comprised of 3, 4, or 5 layered units described above:

\begin{eqnarray} E = \sum_i N_i c_i, 
  \label{eqn:1}
\end{eqnarray}

\begin{figure}[!t] \centering
  \includegraphics[width=0.48\textwidth]{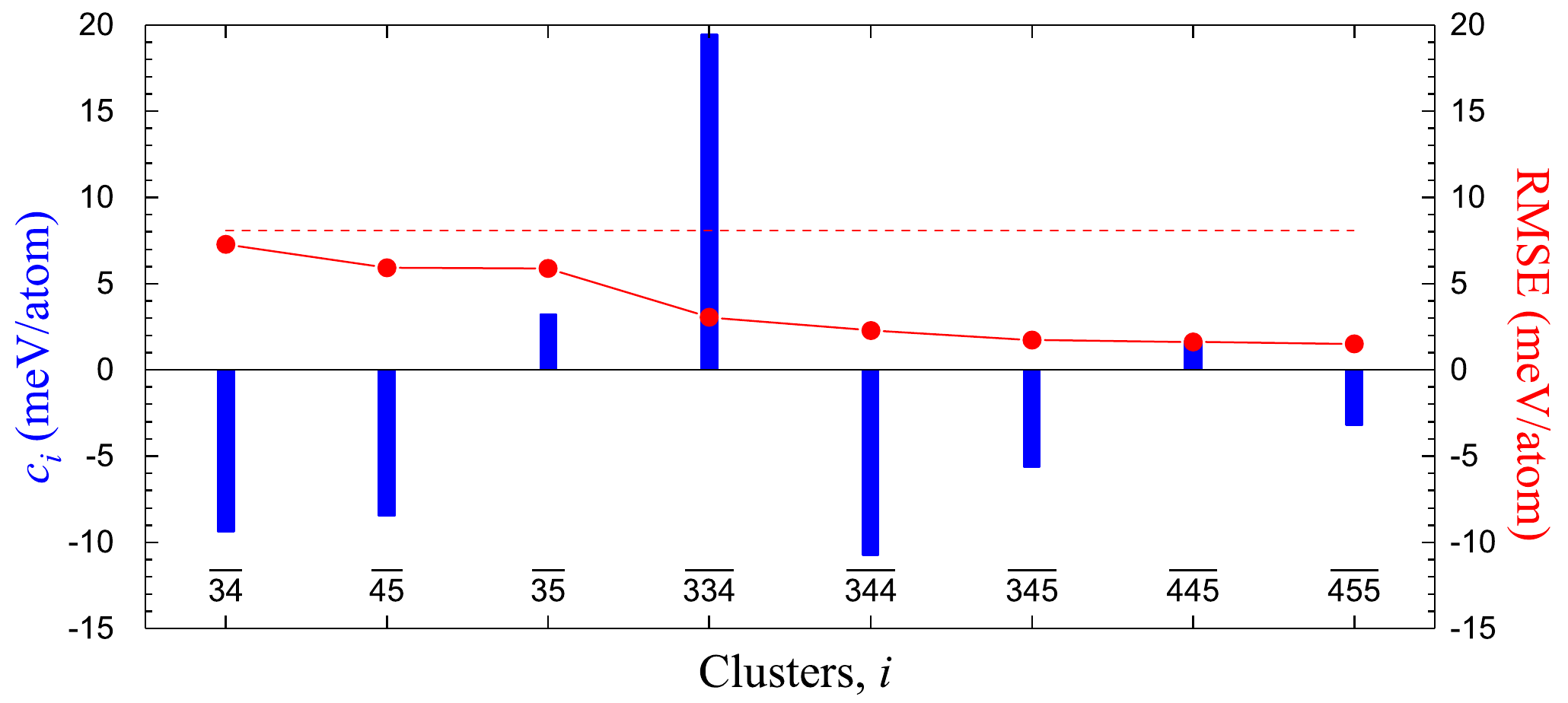} \caption{Strength of
  the expansion coefficients (left vertical axis) and RMSE (right vertical
  axis) in the description of [111] BCC phases using a linear model based on
  blocks with Sn atoms separated by 3, 4, or 5 layers. The $\vert$3453$\vert$,
  phase, for example, has $N_3=N_5=N_{345}=1$ and $N_4=N_{34}=N_{45}=2$ in Eqn.
  \ref{eqn:1} (symmetric combinations, such as $\overline{34}$ and $\overline{43}$, 
  are grouped together). The RMSE
  was calculated for a different number of expansion terms from 3 
  (\{$\overline{3}$,$\overline{4}$,$\overline{5}$\}
  giving 8.1~meV/atom marked with the dashed red line) to 11. The displayed
  coefficient strengths were evaluated for the full set of 11 blocks.}
  \label{fig06:rmse}
\end{figure}
\begin{figure}[!t] \centering
  \includegraphics[width=0.48\textwidth]{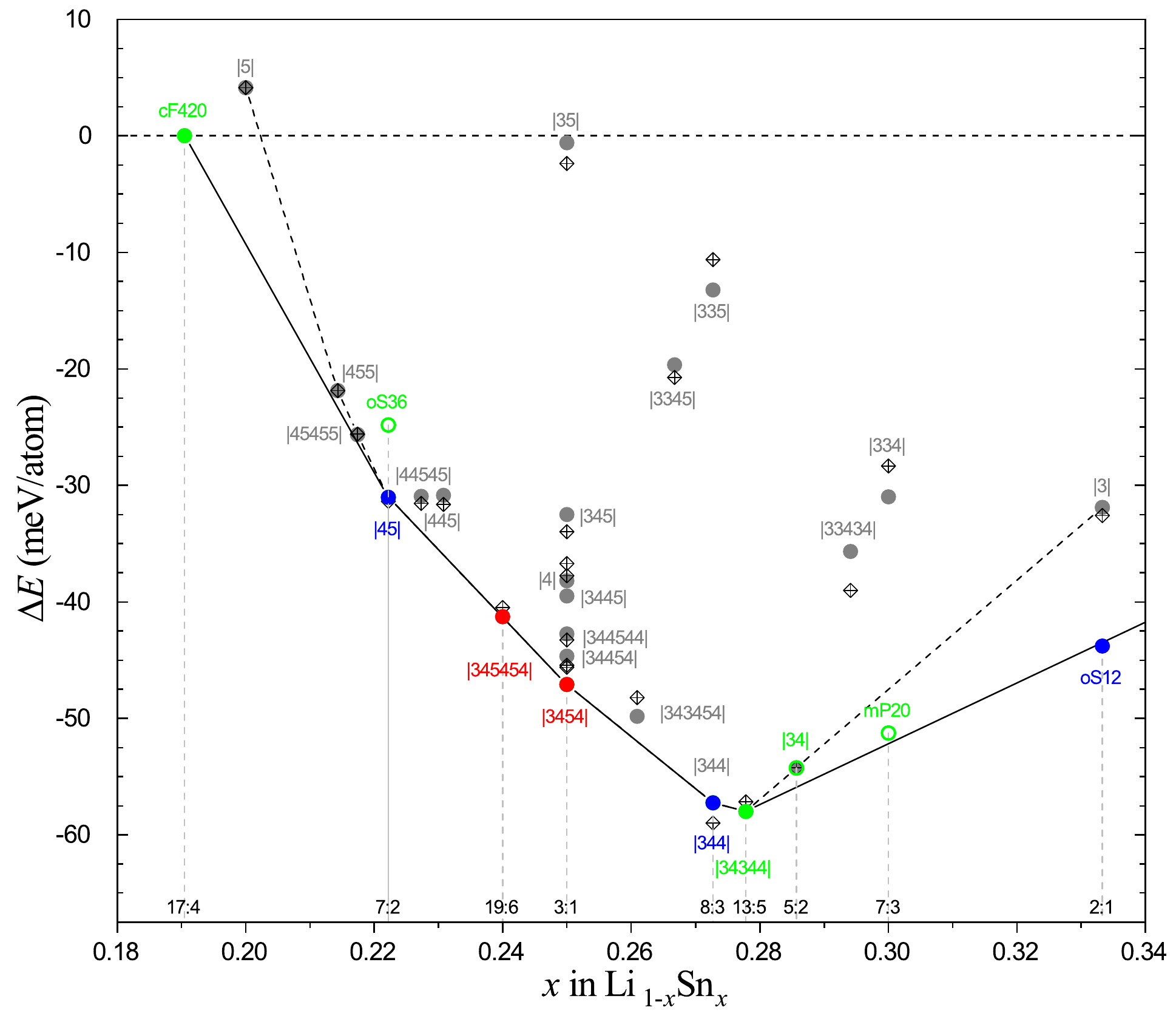} \caption{Stability
  of Li-rich phases relative to cF420-Li$_{17}$Sn$_4$ and mP6-LiSn evaluated
  with the DFT (solid circles).  Stability of phases with the [111] BCC
  morphology, denoted with block sequences explained in the text, is also
  approximated with a phenomenological model (crossed diamonds).  The DFT
  convex hulls shown with dashed and solid lines correspond to the [111] BCC
subset and all considered phases, respectively.  The Li-Sn phases observed
experimentally, proposed previously, and predicted in this study are shown in
green, blue, and red, respectively.  } 
  \label{fig07:bcc}
\end{figure}

\noindent where $c_i$ is the energy of block $i$ and $N_i$ is the number of
such blocks in a structure. Taking advantage of the phenomenological model's
easy interpretability, we considered different blocks, marked with an overbar,
to determine stable unit combinations and create new competitive BCC
structures.  We observed that the trivial single-unit
${\{i\}=\{\overline{3},\overline{4},\overline{5}\}}$ set, providing a linear
interpolation of energies in the 4:1-2:1 composition range, offered a
8.1~meV/atom accuracy comparable to that of the NN model. The final $
{\{i\}=\{\overline{3},\overline{4},\overline{5},\overline{34},\overline{45},\overline{35},\overline{334},\overline{344},\overline{345},\overline{445},\overline{455}\}}$
set approximated the energies with an RMSE of 1.5~meV/atom and showed that the
$\overline{34}$, $\overline{45}$, $\overline{344}$, and $\overline{345}$ blocks
are beneficial while $\overline{35}$ and $\overline{334}$ are detrimental for
structure stability (Fig. \ref{fig06:rmse}). Based on this information, we
constructed a number of new members with up to 6 units and plotted their DFT
relative formation energies (Fig. \ref{fig07:bcc}).

This analysis led to identification of yet another thermodynamically stable BCC
phase, hR75-Li$_{19}$Sn$_6$. Comprised of 6 layered units, the $\vert$345454$\vert$
structure is far more difficult to find with {\it ab initio} searches because
it has an unusual composition and a 25-atom primitive rhombohedral unit cell
with a large $c/a\approx 15$ ratio. Our three NN evolutionary runs with 10$^4$
local relaxations failed to locate the presumed 19:6 global minimum and
converged to a monoclinic mS50 structure 2.1~meV/atom (7.3~meV/atom in the PBE)
above hR75. According to our DFT tests (Table S1), hR75-Li$_{19}$Sn$_6$ lies
essentially on the hP9-Li$_7$Sn$_2$ $\leftrightarrow$ hR48-Li$_3$Sn tie-line and
below the hP9-Li$_7$Sn$_2$ $\leftrightarrow$ hR33-Li$_8$Sn$_3$ tie-line by 
1-2~meV/atom.

Alloys with higher Sn content ($x \geq 0.3$) adopt stable configurations with
distinctive extended Sn fragments or frameworks \cite{Mayo2017}. The observed
mP20-Li$_7$Sn$_3$ phase \cite{muller1974} is a peculiar decoration of the BCC lattice
resulting in Sn zig-zag trimers with relatively short bonds of 2.99 and 3.00
\AA.  It has been found slightly metastable by 1-2~meV/atom with respect to
hP18-Li$_{13}$Sn$_5$ and mP6-LiSn in the standard DFT calculations (Table S1)
\cite{Mayo2017,Sen2017}. Our phonon calculations suggest a stabilizing effect of
the vibrational entropy but the flatness of the relative free energy curve
makes it difficult to give a reliable estimate of the phase transition
temperature (Fig. \ref{fig:hull00}). Moreover, our SCAN calculations show that
it may be a true ${T=0}$~K ground state.

The previously proposed oS12-Li$_2$Sn phase \cite{Mayo2017} was found to be the most
stable 2:1 configuration in our calculations as well. The structure contains
planar zig-zag Sn chains separated by two Li layers. In contrast to other Li-Sn
phases discussed so far, the Li and Sn sites have 12-atom coordinations and are
distributed on an HCP lattice. The PBE calculations with an ultrasoft
pseudopotential placed it just 1~meV/atom above the
hP18-Li$_{13}$Sn$_5$ $\leftrightarrow$ mP6-LiSn tie-line \cite{Mayo2017}. Our PBE and SCAN
results (Table S1) indicate that oS12-Li$_2$Sn may actually be
thermodynamically stable at zero and elevated temperatures (Fig.
\ref{fig:hull00}).

The experimental mP6-LiSn phase \cite{muller1973} is another example
of a stable BCC alloy. A clustered population of the lattice sites
creates fully connected zig-zag Sn nets shown in
Fig. \ref{fig:LiSnStrs}(j). Our evolutionary searches for various unit
cell sizes of the 1:1 composition located several DFT local minima
within a 20~meV/atom energy window (Figs. \ref{fig:hull00} and S1). The most dramatic
entropy-driven stabilization relative to mP6, from 14.9~meV/atom at 0~K
down to 6.4~meV/atom at 600~K, occurred for a tI24 phase observed in
Ref. \onlinecite{tI24}, but none of the considered polymorphs became stable in our
calculations (Figs. S1 and S2) up to the reported melting temperature of
758~K at this composition \cite{natesan2002}.

The most Sn-rich synthesized alloy, tP14-Li$_2$Sn$_5$ with the Hg$_5$Mn$_2$
prototype \cite{hansen1969}, can be viewed as a network of Li-intercalated
pentagonal prisms (Fig. \ref{fig:LiSnStrs}(k)). Located 1.2~meV/atom above the
mP6-LiSn $\leftrightarrow \alpha$-Sn tie-line at ${T=0}$~K, the phase is predicted 
to become stable above 110~K (Fig. \ref{fig:hull00}).

Our screening of competitive high-$T$ compounds produced a viable FCC-based
mS40-LiSn$_4$ phase (Fig. \ref{fig:LiSnStrs}(l)) that breaks the
tP14-Li$_2$Sn$_5$ $\leftrightarrow \alpha$-Sn tie-line at 510~K. This temperature
estimate is particularly sensitive to the description of pure Sn. The
difficulty of reproducing $\alpha \rightarrow \beta$ transition temperatures
and pressures for group-XIV Si and Sn has been pointed out in several studies
\cite{pavone1998, legrain2016, sorella2011, ravelo1997, nagy1999, christensen1993, mehl2021}. The overestimated temperature value for the
$\alpha$-Sn to $\beta$-Sn transformation obtained in our standard PBE
calculations is 540~K. A lower free energy of the reference $\beta$-Sn could
shift the mS40-LiSn$_4$ stabilization to significantly higher temperatures.

To the best of our knowledge, there have been no attempts yet to synthesize
Li-Sn alloys at high pressures, which makes Sen and Johari's {\it ab initio}
study \cite{Sen2017} an important baseline for stability of binary compounds up
to 20~GPa. Our evolutionary searches performed at 20~GPa reproduced all the
previously reported ground states for compositions above $x=2/9$. The 7:2, 8:3,
13:5, and 5:2 compounds retain their ambient-pressure BCC prototypes discussed
above, while 3:1 and 1:1 compounds adopt well-known simpler cF16 ($D0_3$) and
cP2 ($B2$) prototypes, respectively. Compared to the ambient-pressure results,
there are very few competing phases within the 20~meV/atom window above
$x=1/3$. The closest to stability phase is tI12-LiSn$_2$, which is 1~meV/atom
above cP2-LiSn $\leftrightarrow$ BCT-Sn as seen in Fig. \ref{fig:hull20}.

For compositions below $x=2/9$, our searches produced several new stable Li-Sn
phases that reshape the Li-rich part of the previously proposed convex hull at
20~GPa. Given the relatively low symmetry of these phases and the apparent lack
of well-defined underlying motifs, their structural features are better
illustrated with the RDF plots shown in Fig. S3. It can be seen that Sn atoms in
these alloys are separated by at least 4.0 \AA{}, and the Sn local environments
consist of 14-17 Li atoms 2.48-3.06 \AA{} away, with more Sn-rich phases
generally having shorter Sn-Li interatomic distances. Despite the morphological
commonalities, the slight variations lead to noticeable differences in
enthalpies. 

At the 7:1 stoichiometry, we obtained several competing configurations with low
symmetry. The most stable aP32 structure is $-1.9$~meV/atom below aP16
proposed previously \cite{Sen2017}. Both aP16 and aP32 have Sn sites with
well-defined shells of 16 Li atoms within the 2.6-3.0 \AA{} range. By linking
Sn atoms with 4.5-5.0-\AA{} connections, one can discern different extended Sn
frameworks with channels of variable size along $a$, larger in aP32, filled
with Li atoms.

According to our global optimization results, the most Li-rich ground state at
20~GPa actually occurs at the 11:2 composition and has a low-symmetry aP26
structure. It makes all 7:1 phases metastable by at least 3.7~meV/atom. While
both the NN and DFT methods favor the aP26 polymorph, the latter indicates that
another 11:2 phase, mP26, is only 0.4~meV/atom above the putative ground state
at $T=0$~K and becomes stable at 460~K. Notwithstanding the near degeneracy in
the full temperature range, the two structures feature different Li
coordinations of the two Sn sites: 16 and 16 in aP16 versus 15 and 17 in mP26.

The most significant improvement in stability with respect to the previously
proposed ground states is observed at the 5:1 stoichiometry. The best
configurations identified in previous \cite{Sen2017} and our studies have the
same Pearson symbol (mS48) and space group ($C2/m$) but the latter is more
stable by 10.7~meV/atom; in fact, the former is only metastable by 8.9~meV/atom
relative to the new set of ground states at neighboring compositions (Table
S2). Morphologically, the previous phase features only $4i$ Li/Sn Wyckoff
positions, Sn sites coordinated with 15 and 17 Li atoms 2.61-3.06 \AA{} away,
and a Sn framework with Li-filled channels along $b$, whereas ours contains
four $8j$ Li Wyckoff positions, Sn sites coordinated with 14 Li atoms only
2.55-2.75 \AA{} away, and a Sn framework without apparent extended pores.

A mS44-Li$_9$Sn$_2$ phase found in our study represents another stable compound
at a new stoichiometry, being at least $-8$~meV/atom lower in enthalpy with
respect to the best 5:1 and 15:4 phases. Compared to the compressed structures
discussed so far, mS44 has Sn-Li interatomic distances dispersed over a wider
2.50-3.36 \AA{} range. The inclusion of this phase in the convex hull
destabilizes the previously proposed tI50-Li$_4$Sn \cite{Sen2017} by about
1~meV/atom (see Table S2).

Finally, we sampled the unusual 15:4 composition because phases with the
cI76-Cu$_{15}$Si$_4$ prototype ($I\overline{4}3d$) have been observed in
related Li-Si and Li-Ge binaries \cite{kubota2007,johnson1965} and because
another oS152-Li$_{15}$Si$_4$ prototype ($Fdd2$) has been shown to become
stable in the Li-Si system above 7.9~GPa \cite{zeng2015}. Our DFT calculations
at 20~GPa indicate that the cubic structure is unstable by over 60~meV/atom but
the orthorhombic one is actually $-$1.3~meV/atom below the tie-line connecting
previously proposed 5:1 and 7:2 ground states \cite{Sen2017} and could have been
considered stable. We find that oS152-Li$_{15}$Sn$_4$ breaks the new 9:2 and
7:2 tie-line as well, albeit by only $-$0.6~meV/atom. However, our evolutionary
searches identified a new hR57 structure $-$1.8~meV/atom lower in enthalpy
compared to oS152. This presumed $T=0$~K ground state with a large rhombohedral
unit cell belongs to the [111] BCC family and can be represented simply as
$\vert4555\vert$. Our phonon calculations demonstrate that oS152 does stabilize over hR57
above 400~K.

The impressive number of viable phases with large low-symmetry unit cells
appearing at ambient and high pressures testifies to the exceptional complexity
of the Li-Sn phase diagram. While we find it encouraging that our extensive
screening accelerated with machine learning uncovered several new phases with
lower Gibbs free energies, it cannot be considered exhaustive. Future
explorations of the Li-Sn system with higher-accuracy MLPs and DFT
approximations may lead to further revision of the phase diagram.

\subsection{Analysis of electronic and other properties of ambient-pressure
Li-Sn compounds}

Electronic, mechanical, and electrochemical properties of Li-Sn alloys have
been investigated in several {\it ab initio} studies
\cite{stournara2012,zhang2015,Mayo2017, Sen2017, Li2013}. Analyses of the
projected DOS and Bader decomposition data have indicated the expected
metallicity of all Li-Sn alloys and transfer of charge from Li to Sn.
Interestingly, the charge redistribution in alloys with high Li concentrations
exceeding 4:1 has been shown to leave Li atoms in either ionized (+0.8) or
nearly neutral ($-$0.3 - 0.0) states \cite{Sen2017}. The manifold of the
electronic states near the Fermi level has primarily the Sn-$p$ character but
displays increasing Li-$s$/$p$ contributions in Li-rich compounds
\cite{zhang2015, Sen2017}. It has been pointed out that addition of Li makes
the dominant bonding in Li-Sn alloys evolve from covalent Sn-Sn/Li-Sn with a
characteristic pseudo-gap to metallic Li-Li \cite{zhang2015}.  

Examination of DOS profiles near the Fermi level can indeed provide insights
into materials’ various properties ranging from transport, superconductivity,
and magnetism to stability. The presence of a pseudo-gap and the population of
bonding states has been found to correlate with structure’s favorability in
numerous cases \cite{ak09, ak31, ak44}. For instance, our studies of related chemical
systems have shown that the unusual stoichiometry of LiB$_x$ ($x \sim 0.9$) and the
stability of NaSn$_2$ result from the placement of the Fermi level near the
bottom of a well-defined DOS valley \cite{ak09, ak30, ak31}.  

\begin{figure}[t!] \centering
  \includegraphics[width=0.48\textwidth]{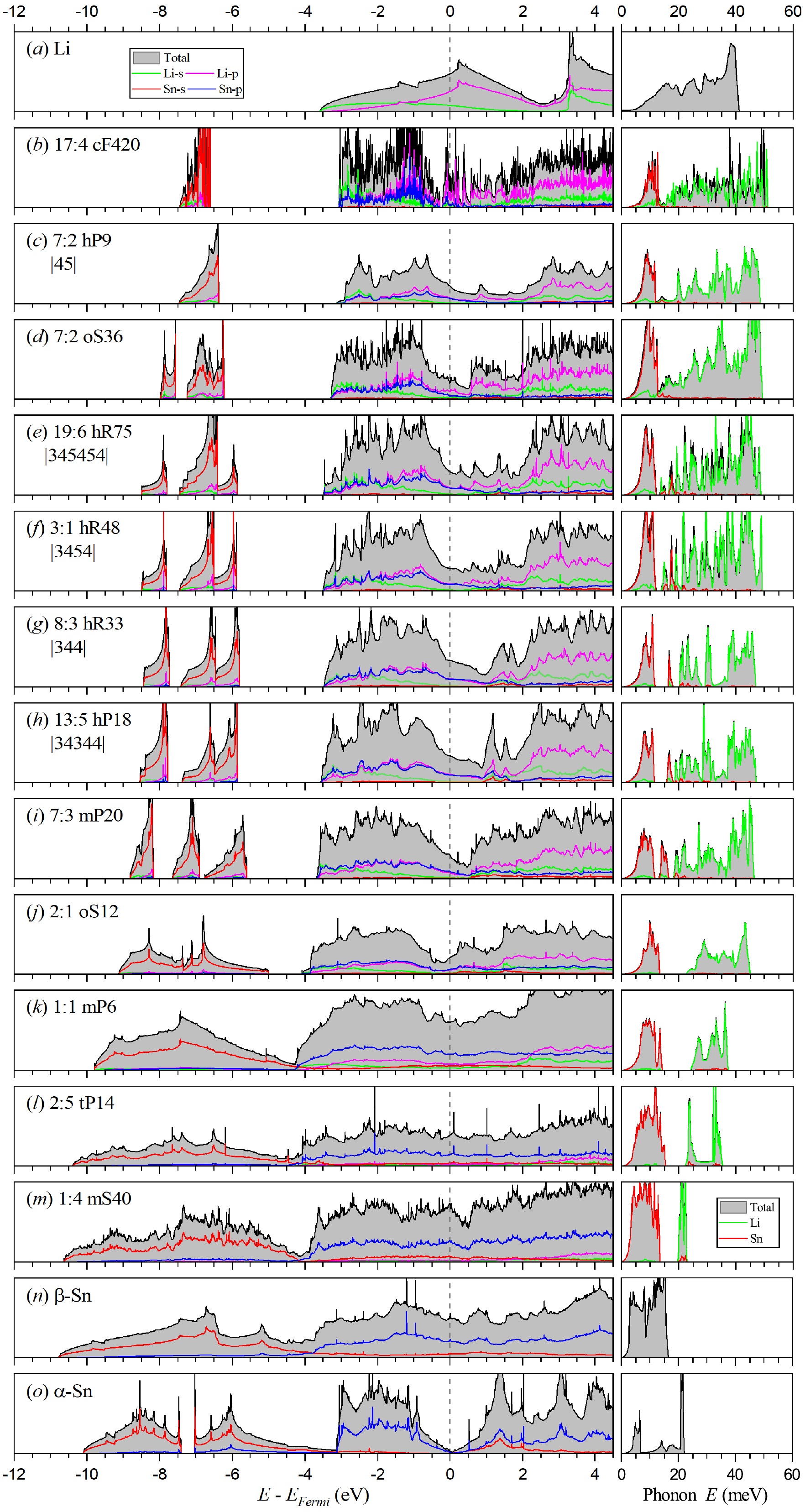} 
  \caption{ Electron (left) and phonon (right) DOS calculated with the DFT for
  known and predicted ambient-pressure Li-Sn phases. The colored lines
  illustrate the DOS projected onto $s$ and $p$ orbitals for electrons or onto
  Li and Sn atoms for phonons. 
  }
  \label{fig:LiSnDOS}
\end{figure}

Figure \ref{fig:LiSnDOS} shows that most of the known and proposed Li-Sn alloys,
especially those with $x \geq 0.5$, have a significant DOS at the Fermi level and
may not conform to this simple model. In fact, it is not easy to disentangle
the hybridization and charge transfer factors that determine the shape and the
population of the electronic states with the Sn-$p$ and Li-$s$/$p$ character. The
Sn-$s$ states, on the other hand, exhibit little hybridization and provide
information about the behavior of Sn-centered orbitals. With the large $s$-$p$
splitting disfavoring the formation of the traditional $sp^2$/$sp^3$ hybrids in
pure Sn open structures, the low-lying $s$ states split first and only then does
the antibonding $s$ set mixes partially with the $p$ states (Fig.
\ref{fig:LiSnDOS}(o)) \cite{ak33, Xu2013}.  

In the $\beta$-Sn, mS40-LiSn$_4$, tP14-Li$_2$Sn$_5$, and mP6-LiSn phases, the
Sn-$s$ states disperse into a single band that partially overlaps in energy
with the bottom of the Sn-$p$ band. As the number of Sn nearest atoms within
the 3.08-3.40 \AA{} range changes from 6 to 8-12, 6, and 4 in this series, the
Sn-$s$ bandwidth gradually decreases. In oS12-Li$_2$Sn with zig-zag chains, the
dispersion reduces down to about 4 eV which creates a 1-eV separation between
the Sn-$s$ set and the Sn-$p$ and Li-$s$/$p$ manifold. All shown phases with
higher Li content do not have extended Sn frameworks and feature fairly
localized Sn-$s$ states. The number and the width of the DOS peaks are
determined by the distribution of isolated Sn trimers (in mP20-Li$_7$Sn$_3$),
Sn dimers (for BCC phases with $0.24 < x < 0.277$), and single Sn atoms (in
hP9-Li$_7$Sn$_2$ and cF420-Li$_{17}$Sn$_4$).  For example, the presence of the
2.9-\AA{} Sn-Sn dimers within the type-3 units in the [111]-BCC
phases (Fig \ref{fig:LiSnStrs}(e-h)) produces two 0.7-eV peaks centered about
~2 eV apart, while the orbital overlaps between Sn atoms ~4.7 \AA{} apart
within the [111]-BCC base or the type-4 blocks give rise to the
central 1-eV peaks (Fig \ref{fig:LiSnDOS}(c,e-h)). The lower edge of the Sn-$p$
band moves in concert with the bandwidths of the Sn-$s$ states, shifting from
$-$4 eV in oS12-Li$_2$Sn to $-$3 eV in cF420-Li$_{17}$Sn$_4$.  Interestingly, the
lowest states in the manifold in oS12-Li$_2$Sn and hP9-Li$_7$Sn$_2$ have the
Li-$s$/$p$ character due to a larger dispersion of these nearly free electron
states that can be seen more clearly on band structure plots for select phases 
(Figs. S4-S6). 

\begin{figure}[t!] \centering
  \includegraphics[width=0.48\textwidth]{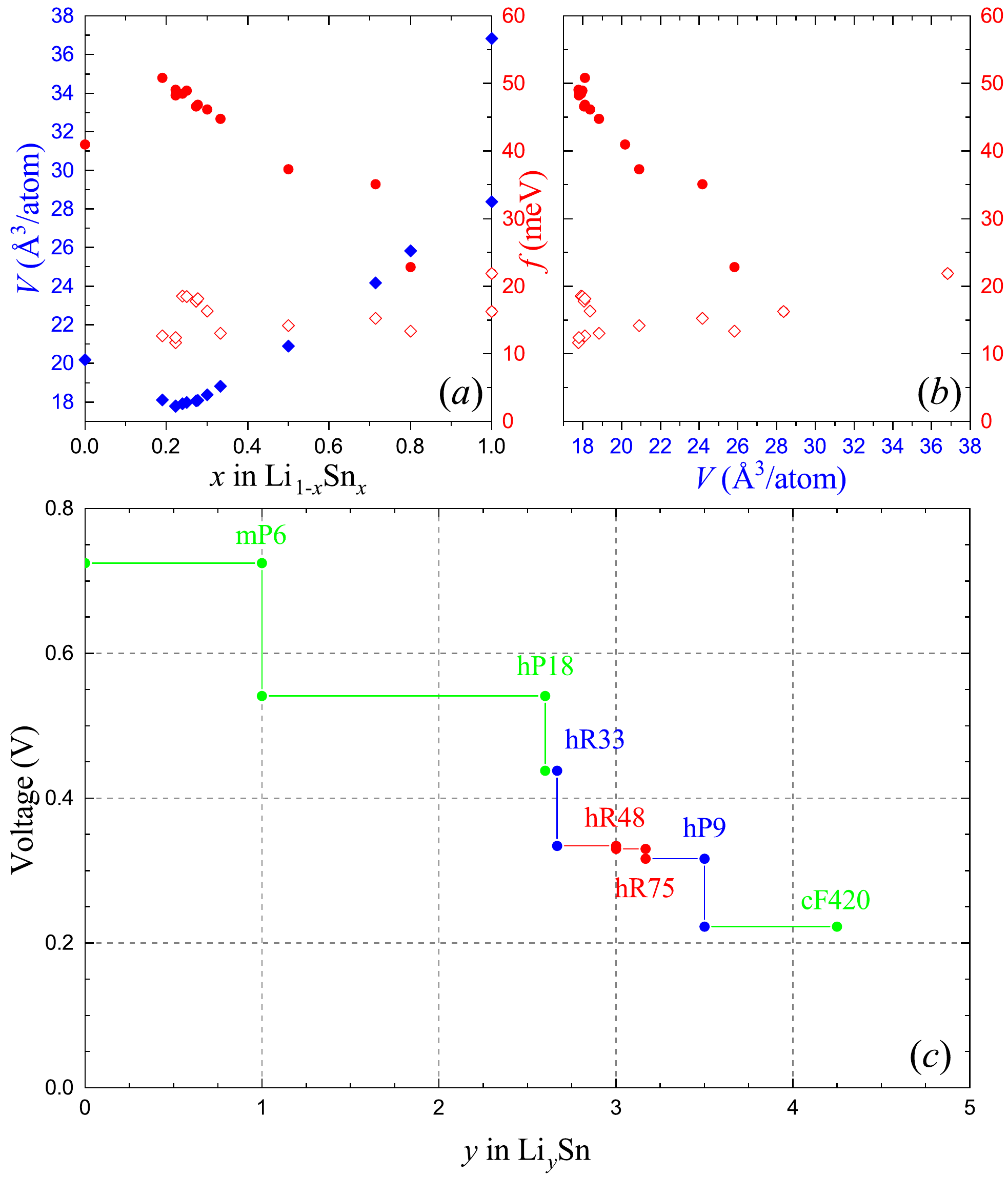} 
  \caption{ (a) Atomic volume (blue diamonds), highest Li phonon frequencies
  (solid red circles), and highest Sn phonon frequencies (empty red diamonds)
  evaluated with the DFT at ambient pressure as a function of the alloy
  composition. (b) The same highest Li and Sn phonon frequencies as a function
  of the atomic volume. (c) Voltage for stable Li$_y$Sn alloys refined through
  addition of two hR48-Li$_3$Sn and hR75-Li$_{19}$Sn$_6$ ground states
  predicted in this study.
  }
\label{fig:LiSnVoltFreq} \end{figure}

The change in volume upon Li insertion is one of the critical factors for
durability of Li-ion battery anodes. It has been shown that the calculated
atomic volume in Li-Sn alloys notably deviates from the linear dependence on
the Li content \cite{Li2013, Sen2017}. Our results in Fig.~
\ref{fig:LiSnVoltFreq} demonstrate that the atomic volume correlates with the
formation energies across the Li$_{1-x}$Sn$_x$ composition range. The lowest
values occur in compounds with the maximum transfer of charge from Li to Sn
near the 4:1 stoichiometry.

A similar trend in the full composition range can be observed for vibrational
properties using the phonon DOS results in Fig. \ref{fig:LiSnDOS}.  The highest
phonon frequencies of 41~meV in BCC-Li, 22~meV in $\alpha$-Sn, and 16~meV in
$\beta$-Sn indicate that the force constants are higher in Sn than in Li
because the mass factor would scale down the Sn frequencies by a larger
$\sqrt{m_{Sn}/m_{Li}}=4.1$. The top edge of the Sn DOS peak downshifts to
13~meV in FCC-based mS40-LiSn$_4$, goes back up to 15~meV in tP14-Li$_2$Sn$_5$
based on a 3D Sn framework, and eventually settles at ~12~meV in BCC-based
Li-rich alloys. The highest frequencies of the Li modes actually increase with
the first addition of Sn, reaching a maximum of 51~meV in
cF420-Li$_{17}$Sn$_4$, and then drop precipitously in alloys with $x \geq 0.5$.
Plotted versus the atomic volume in Fig. \ref{fig:LiSnVoltFreq}(b), the highest
phonon frequencies in all Li-Sn phases exhibit a clearer trend: the BCC-Li and
$\alpha$-Sn outliers fall closer to the linear fit, which leaves
tP14-Li$_2$Sn$_5$ with hard Li modes as the only significant deviation.  

Finally, we comment on how the predicted ambient-pressure phases could be
detected. Electrochemical cycling measurements have been used to identify Li-Sn
alloys in several studies \cite{Wang1986, Courtney1998, Tran2011}. We update the
previously reported average voltage plot \cite{Mayo2017} by including the
hR75-Li$_{19}$Sn$_6$ and hR48-Li$_3$Sn phases predicted in our study. Since
these phases deepen the convex hull by only 1-2~meV/atom, the additional
features near $y=3$ in Li$_y$Sn (Fig. \ref{fig:LiSnVoltFreq}) are very fine and
would be difficult to observe.  However, their powder x-ray diffraction (XRD) patterns differ
considerably from those in previously observed or proposed BCC phases and would
provide strong evidence of their formation (Fig. S7). The combination of
electrochemical and XRD measurements have been recently used to detect a new
NaSn$_2$ phase \cite{Stratford2017} predicted in our earlier {\it ab initio}
study \cite{ak31}.

\section{Summary} Our re-examination of the Li-Sn binary illustrates that
well-known systems may still host new synthesizable compounds and that
computationally demanding {\it ab initio} prediction of thermodynamically stable
materials can be effectively boosted with MLPs. Given the surprising scarcity
of successful MLP-assisted predictions of stable compounds so far, our study
was focused on devising guidelines for constructing reliable models,
identifying viable structures, and establishing stability trends that increase
the chance of finding {\it ab initio} ground states. To this extent, the combination
of the evolutionary sampling and stratified training allowed us to build a
practical Li-Sn NN potential with a relatively modest 10.2~meV/atom accuracy
but a reliable performance in unconstrained searches. The model construction
involved less than 2\% of the total computational cost required for the
systematic exploration of the binary system. Our screening of over 1.1 million
(2.3 thousand) crystalline phases at zero (elevated) temperature with the NN
potential along with the following DFT analysis of select candidates cost only
an estimated 0.1-1\% of the traditional {\it ab initio} searches. As a result, we
discovered overlooked thermodynamically stable hR48-Li$_3$Sn and mS40-LiSn$_4$
at ambient pressure and several possible new ground states at 20~GPa. In
contrast to the previously identified stable hP9-Li$_7$Sn$_2$ and
hR33-Li$_8$Sn$_3$ phases \cite{Sen2017, Mayo2017} that have known Li$_7$Ge$_2$
and Li$_8$Pb$_3$ prototypes in related binaries and can be naturally identified
by chemical substitution, our large hR48 and mS40 structures appear to have no
analogues in other chemical systems. Moreover, after introducing a
phenomenological model for the [111] BCC family of Li-rich alloys, we
constructed another stable hR75-Li$_{19}$Sn$_6$ phase with an even larger unit
cell that could not be found with our standard evolutionary searches.  These
findings highlight the benefit of combining different prediction strategies
that utilize available experimental knowledge, explore configuration spaces
with global optimization algorithms, and involve rational design based on
stability analysis.  Since the thermodynamic stability of the new compounds
predicted with this method is ultimately determined by the accuracy of the
employed DFT approximation, we performed additional tests with various DFT
flavors. Our results indicate that the relatively small formation enthalpy
differences of a few meV/atom are fairly insensitive to the systematic and
numerical errors due to the similarity of the competing configurations. For the
same reason, the vibrational entropy contribution to the Gibbs free energy was
found to be an insignificant factor for differentiating Li-rich phases with
related BCC motifs. Consequently, the vast majority of detected metastable
phases in the 20~meV/atom enthalpy window, a common criterion for viable
candidates, proved to be irrelevant in the Li-Sn system. The proposed
thermodynamically stable phases are expected to be synthesizable and detectable
with standard powder XRD measurements. The formation of hR48-Li$_3$Sn and
hR75-Li$_{19}$Sn$_6$ would lead to further refinement of the electrochemical
profile at the Li-rich end of the complex phase diagram.\\

\section*{Acknowledgements} 
We acknowledge the NSF support (Award No. DMR-1821815) and the Extreme Science
and Engineering Discovery Environment computational resources \cite{exsede} (NSF Award
No. ACI-1548562, Project No. TG- PHY190024).  

\section*{Competing interests}
The authors declare no competing interests.

\bibliographystyle{apsrev4-2}
\bibliography{lit}

\end{document}


\title{Supplementary Material: Prediction of stable Li-Sn compounds: boosting \emph{ab initio} searches with neural network potentials.} 
\author{Saba Kharabadze, Aidan Thorn, Ekaterina A. Koulakova, and Aleksey N. Kolmogorov}

\affiliation{Department of Physics, Applied Physics and Astronomy, Binghamton University, State University of New York,
PO Box 6000, Binghamton, New York 13902-6000, USA}

\date{\today}

\maketitle



\begin{widetext}
\begin{center}
 \begin{table*}[ht!]
 \hfill{}
\begin{tabular}{p{1cm}p{14cm}p{4cm}}

  I  &  Table \ref{tabS1} Relative stability of select Li-Sn phases at $P=0$ GPa and $T=0$ K  \dotfill     & 2 \\
    & & \\
  II  &  Table \ref{tabS2} Relative stability of select Li-Sn phases at $P=20$ GPa and $T=0$ K \dotfill     & 2 \\
    & & \\
  III  &  Figure \ref{figS1} Stability reordering for Li$_{17}$Sn$_4$, LiSn, and Li$_2$Sn$_5$ at $P=0$ GPa and $T=600$ K \dotfill     & 3 \\
    & & \\
  IV  &  Figure \ref{figS3} Relative free energies of select Li-Sn phases at $P=0$ GPa and $T=0-800$ K \dotfill     & 4 \\
    & & \\
  V  &  Figure \ref{figS4} Radial distribution functions of Li-rich phases at $P=20$ GPa \dotfill     & 5 \\
    & & \\
  VI  &  Figure \ref{figS5} Band structure and density of states in hP9-Li$_7$Sn$_2$ at $P=0$ GPa \dotfill     & 6 \\
  & & \\
  VII  &  Figure \ref{figS6} Band structure and density of states in hR75-Li$_{19}$Sn$_6$ at $P=0$ GPa  \dotfill     & 6 \\
    & & \\
  VIII  &  Figure \ref{figS7} Band structure and density of states in hR48-Li$_3$Sn at $P=0$ GPa   \dotfill     & 7 \\
  & & \\

  IX  &  Figure \ref{figS2} Simulated powder x-ray diffraction patterns for select ambient-pressure phases  \dotfill     & 7 \\
    & & \\

    \end{tabular}
\hfill{}
\end{table*}
\end{center}
\end{widetext}

\newpage

\begin{table}[t]
  \begin{tabular}{D{-}{-}{-1} l r r r r}
\hline
\hline
    \multicolumn{1}{c}{\text{Phase}} &  \multicolumn{1}{c}{\text{Relative to}} & \multicolumn{1}{c}{\text{PBE}} & \multicolumn{1}{c}{\text{PBE}} & \multicolumn{1}{c}{\text{LDA}}   & \multicolumn{1}{c}{\text{SCAN}} \\ 
                                       &        & \multicolumn{1}{c}{\text{500 eV}}   &   \multicolumn{1}{c}{\text{700 eV}}   & \multicolumn{1}{c}{\text{700 eV}} &  \multicolumn{1}{c}{\text{700 eV}} \\ \hline  
    \text{oS36} - \text{Li$_7$Sn$_2$}  &  hP9$-$Li$_7$Sn$_2$                                       & 6.2             & 6.2               & 5.8           & 3.1  \\        	\text{hR75} - \text{Li$_{19}$Sn$_6$}\ast &  hP9$-$Li$_7$Sn$_2$ $\leftrightarrow$ hR33$-$Li$_8$Sn$_3$     & -1.2            & -1.2              & -1.4          & -1.8  \\ 
    \text{hR75} - \text{Li$_{19}$Sn$_6$}\ast &  hP9$-$Li$_7$Sn$_2$ $\leftrightarrow$ hR48$-$Li$_3$Sn$\ast$         & -0.2            & -0.2              & -0.1          &  -0.1  \\ 
    \text{hR48} - \text{Li$_3$Sn}\ast      &  hP9$-$Li$_7$Sn$_2$ $\leftrightarrow$ hR33$-$Li$_8$Sn$_3$     & -1.7            & -1.7              & -2.1          & -2.6  \\ 
    \text{hP18} - \text{Li$_{13}$Sn$_5$} & hR33$-$Li$_8$Sn$_3$ $\leftrightarrow$ mP6$-$LiSn             & -2.3            & -2.3              & -3.1          &  -3.4  \\  
    \text{hR21} - \text{Li$_5$Sn$_2$}  &  hP18$-$Li$_{13}$Sn$_5$ $\leftrightarrow$ mP6$-$LiSn          &  1.8            & 1.8               &  0.5          &  -0.4  \\
    \text{oS12} - \text{Li$_2$Sn}      &  hP18$-$Li$_{13}$Sn$_5$ $\leftrightarrow$ mP6$-$LiSn          & -0.3            & -0.3              & 0.6           &  -3.4  \\ 
    \text{oS32} - \text{LiSn}          &  mP6$-$LiSn                                               & 12.1            & 12.6              & 14.4          &  9.5   \\
\hline
\hline
\end{tabular} \caption {Relative stability of select Li-Sn phases at $P=0$ GPa
  and $T=0$ K. The energy differences (in meV/atom) are calculated with
  different DFT approximations (PBE, LDA, or SCAN) and plane-wave energy
  cutoffs (500 eV or 700 eV) relative to a competing phase at the same
  composition or the tie-line connecting two neighboring ground states.
  Thermodynamically stable phases predicted in this study are marked with
  asterisks.} 
\label{tabS1} 
\end{table}

\begin{table}[h!]
  \begin{tabular}{D{-}{-}{-1} l r r r r }
\hline
\hline
    \multicolumn{1}{c}{\text{Phase}}  & \multicolumn{1}{c}{Relative to} & \multicolumn{1}{c}{PBE} & \multicolumn{1}{c}{PBE} & \multicolumn{1}{c}{LDA} & \multicolumn{1}{c}{SCAN}  \\ 
                       &               &  \multicolumn{1}{c}{500 eV} & \multicolumn{1}{c}{700 eV} & \multicolumn{1}{c}{700 eV} & \multicolumn{1}{c}{700 eV} \\ \hline 
    \text{aP32} - \text{Li$_7$Sn}         & aP16$-$Li$_7$Sn                                                 & -1.9     &      -1.9      &      -1.4     &   -1.4  \\
    \text{mS48} - \text{Li$_5$Sn}         & mS48$-$Li$_5$Sn$\ast$                                                 & 10.7     &     10.7      &      10.4     &   7.5   \\
    \text{mS48} - \text{Li$_5$Sn}         & aP26$-$Li$_{11}$Sn$_2$$\ast$  $\leftrightarrow$ mS44$-$Li$_9$Sn$_2$$\ast$         &  8.9     &       8.9      &       8.7     &  7.1  \\   
    \text{mS44} - \text{Li$_9$Sn$_2$}\ast     & mS48$-$Li$_5$Sn$\ast$     $\leftrightarrow$ hR57$-$Li$_{15}$Sn$_4$$\ast$        & -8.4     &      -8.4      &      -8.3     &   -7.1 \\
    \text{tI50} - \text{Li$_4$Sn}         & mS44$-$Li$_9$Sn$_2$$\ast$  $\leftrightarrow$  hR57$-$Li$_{15}$Sn$_4$$\ast$  &   1.1    &       1.1      &       1.2     &       2.1 \\
    \text{hR57} - \text{Li$_{15}$Sn$_4$}\ast  & cI76$-$Li$_{15}$Sn$_4$                                          & -63.8    &    -63.8      &   -65.5     &   -65.7  \\
    \text{hR57} - \text{Li$_{15}$Sn$_4$}\ast  & oS152$-$Li$_{15}$Sn$_4$                                          & -1.8     &      -1.8      &      -1.6     &    -2.0  \\
    \text{hR57} - \text{Li$_{15}$Sn$_4$}\ast  & mS44$-$Li$_9$Sn$_2$$\ast$  $\leftrightarrow$  hP9$-$Li$_7$Sn$_2$      &  -2.3    &      -2.3      &      -3.4     &      -3.6 \\
    \text{oS152}- \text{Li$_{15}$Sn$_4$}  & mS44$-$Li$_9$Sn$_2$$\ast$  $\leftrightarrow$   hP9$-$Li$_7$Sn$_2$     &  -0.6    &      -0.6      &      -1.8     &      -1.6 \\
    \text{hP9}  - \text{Li$_7$Sn$_2$}     & hR57$-$Li$_{15}$Sn$_4$$\ast$ $\leftrightarrow$ cF16$-$Li$_3$Sn              & -7.4     &      -7.4      &      -7.2     &   -7.1     \\
    \text{cF16} - \text{Li$_3$Sn}         &  hP9$-$Li$_7$Sn$_2$  $\leftrightarrow$ hR33$-$Li$_8$Sn$_3$            & -3.2     &      -3.3      &      -3.9     &   -1.8  \\
    \text{hR33} - \text{Li$_8$Sn$_3$}     & cF16$-$Li$_3$Sn   $\leftrightarrow$ hP18$-$Li$_{13}$Sn$_5$              & -2.9     &      -2.9      &      -2.8     &   -3.2  \\
    \text{hP18} - \text{Li$_{13}$Sn$_5$}  & hR33$-$Li$_8$Sn$_3$  $\leftrightarrow$ hR21$-$Li$_5$Sn$_2$            & -1.3     &      -1.3      &      -1.2     &   -1.5  \\
    \text{hR21} - \text{Li$_5$Sn$_2$}     & hP18$-$Li$_{13}$Sn$_5$ $\leftrightarrow$ cP2$-$LiSn             & -0.5     &      -0.5      &      -2.0     &   -0.8  \\

\hline
\hline
\end{tabular}
  \caption {Relative enthalpy (in meV/atom) for select Li-Sn phases at $P=20$ GPa and $T=0$ K. }
  \label{tabS2}
\end{table}

\normalsize
\newpage

\newpage

\begin{figure}[ht!]
\centering
  \includegraphics[width=0.9\textwidth]{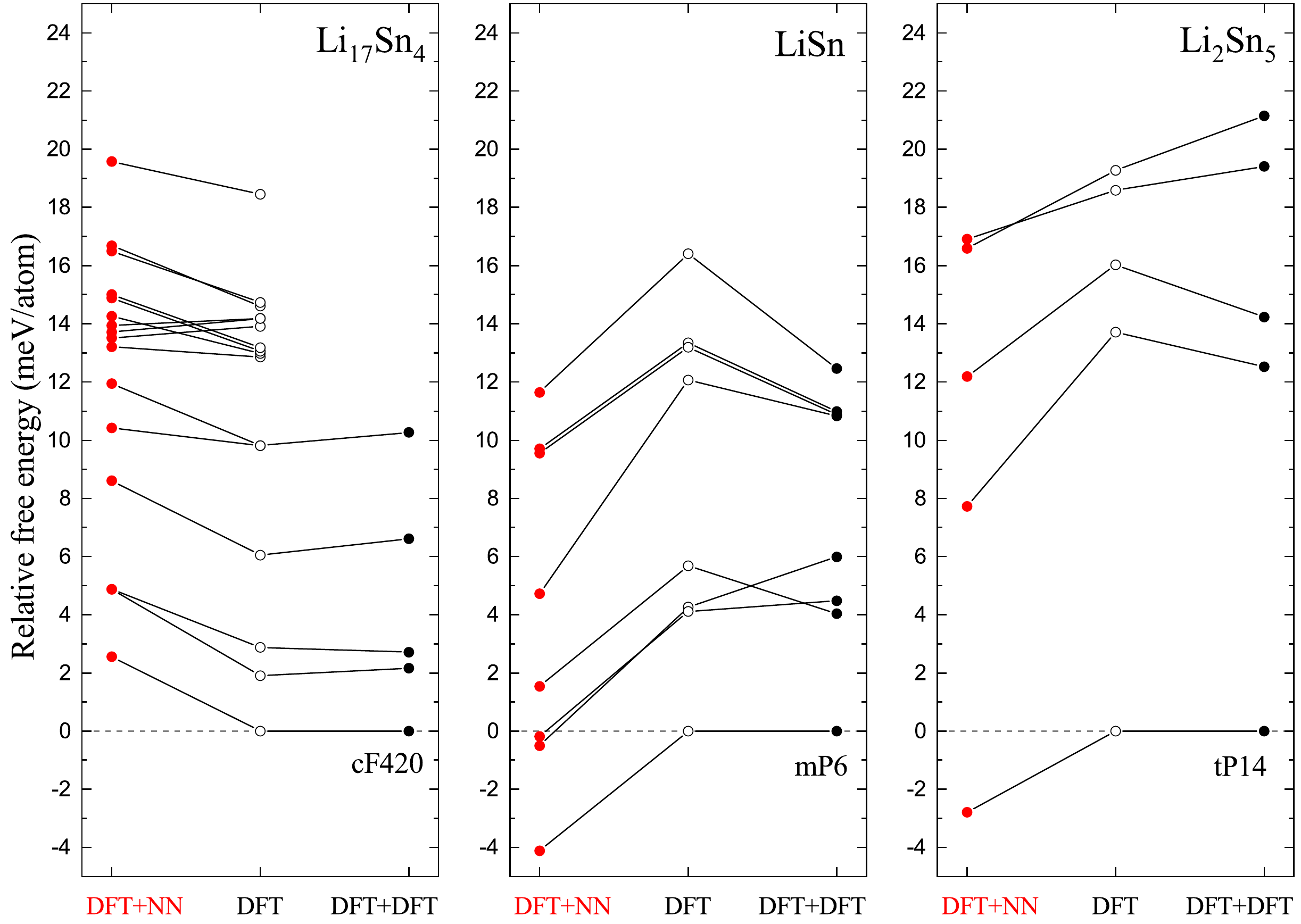}
  \caption{ Examples of stability reordering for competing ambient-pressure
  phases at 600 K after the inclusion of the vibrational entropy contribution
  to the free energy evaluated at the NN and DFT levels. In each panel, the
  hollow black points in the middle (marked as ‘DFT’) show the DFT relative
  energies (without zero point corrections) of candidate structures found at
  the specified composition with our NN-based evolutionary searches; the solid
  red points on the left (‘DFT+NN’) show the relative free energies obtained by
  adding the NN thermal corrections to the DFT energies; and the solid black
  points on the right (‘DFT+DFT’) illustrate the relative free energies
  obtained by adding the DFT thermal corrections. To demonstrate the
  correspondence between the ‘DFT+NN’ and ‘DFT+DFT’ results, both sets are
  referenced to the ‘DFT+DFT’ free energy of the best candidate structure. Due
  to the cost of the DFT phonon calculations, only a few candidates with the
  lowest ‘DFT+NN’ free energies were re-evaluated at the DFT level at each
  composition.  } 
\label{figS1} 
\end{figure}

\begin{figure}[!ht]
\centering
  \includegraphics[width=0.9\textwidth]{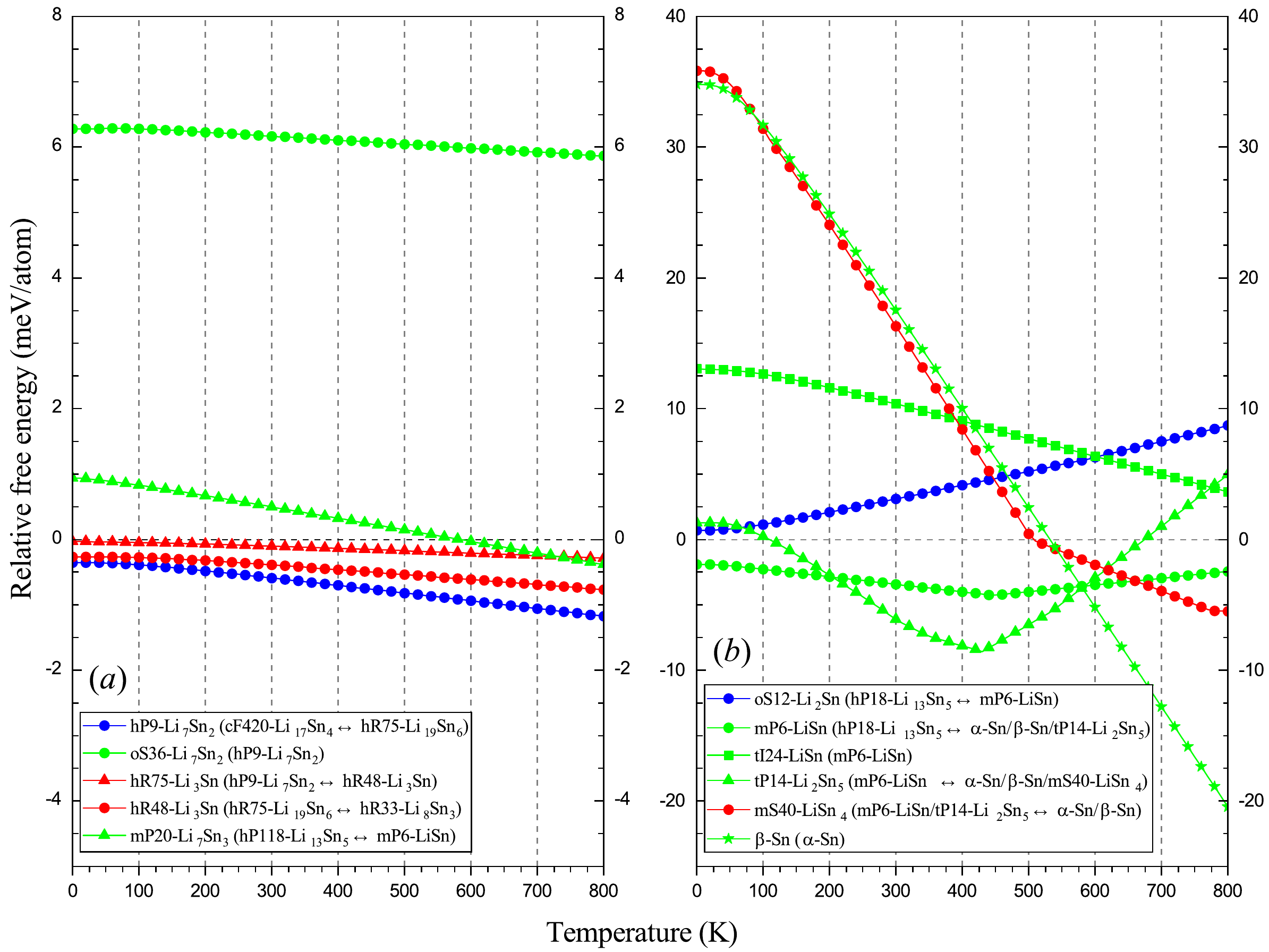} \caption{DFT free energies
  for select ambient-pressure Li$_{1-x}$Sn$_x$ alloys relative to the best
  same-composition phase or mixture of phases (shown in parentheses) at each
  temperature in the 0-800 K range (the values at ${T=0}$~K include zero point 
  energies). Negative free energy values correspond to
  stable phases. Panels (a) and (b) show Li-rich phases ($x<0.3$) and Sn-rich
  phases ($x\geq0.3$), respectively.} 
\label{figS3} 
\end{figure}

\begin{figure}[!ht]
\centering
  \includegraphics[width=0.85\textwidth]{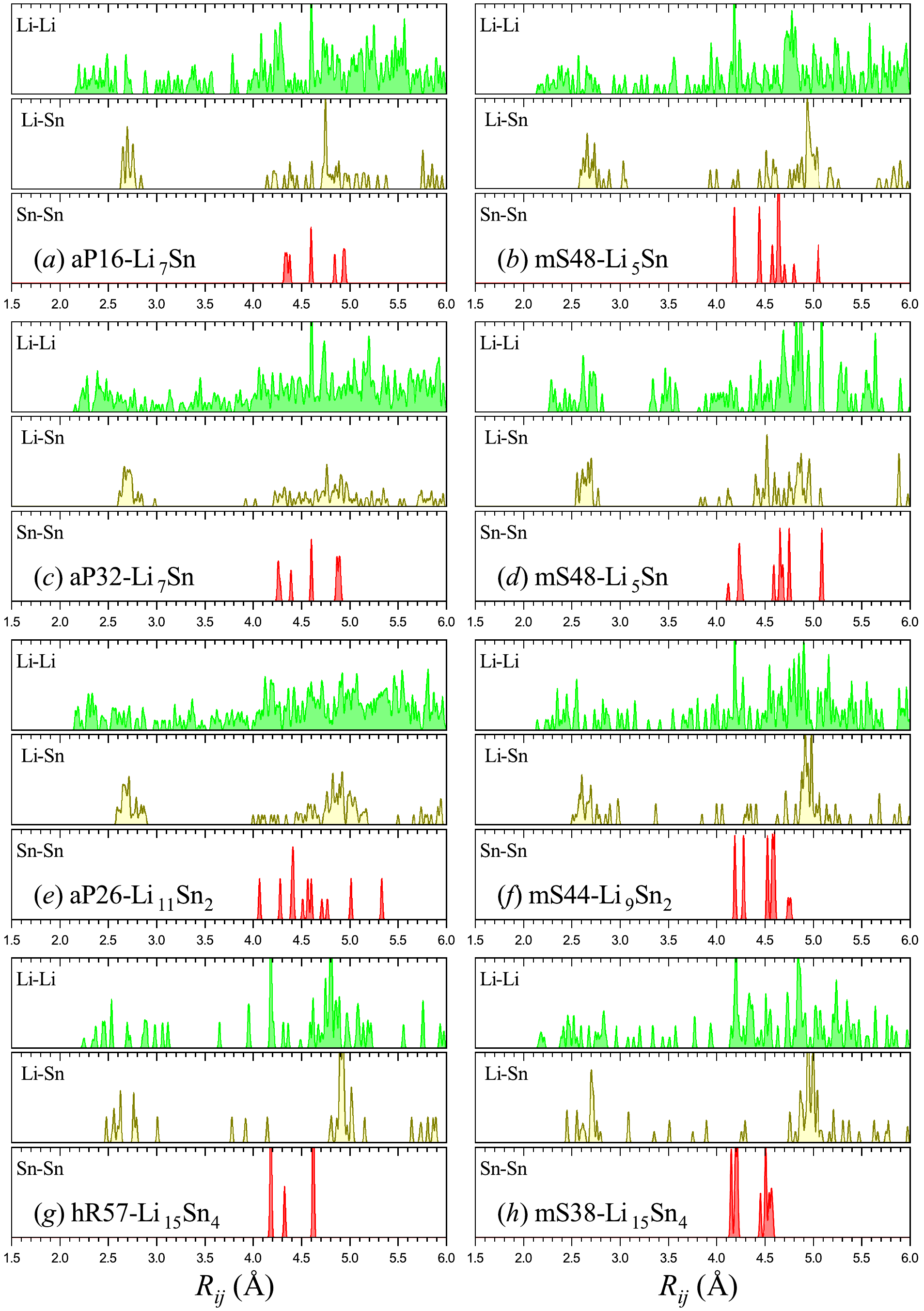}
  \caption{ Radial distribution functions for Li-Li, Li-Sn, and Sn-Sn pairs in
  select phases fully optimized with the DFT at $P=20$ GPa. A
  $\sigma=0.008$-\AA\ Gaussian smearing was used to make the histograms.
  }
  \label{figS4}
\end{figure}

\begin{figure}[ht]
\centering
  \includegraphics[width=0.8\textwidth]{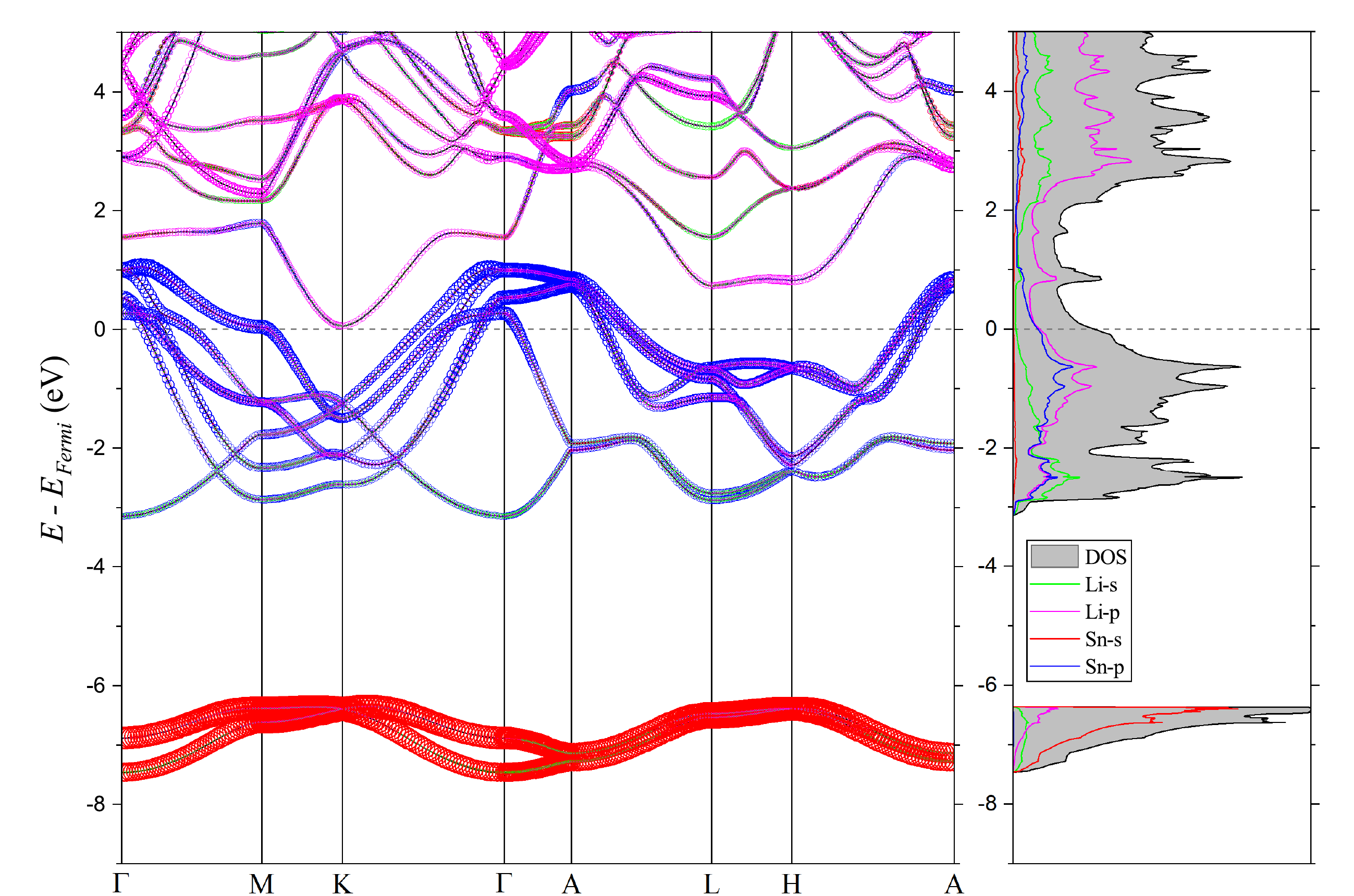} \caption{Band structure and
  density of states (DOS) for hP9-Li$_7$Sn$_2$ at $P=0$ GPa. The strength of
  the electronic states’ orbital character is proportional to the circles size
  in the left panel and the value of the projected DOS in the right panel,
  while the type of the orbital character is displayed with colors specified in
  the legend.} 
\label{figS5} 
\end{figure}

\begin{figure}[ht]
\centering
  \includegraphics[width=0.8\textwidth]{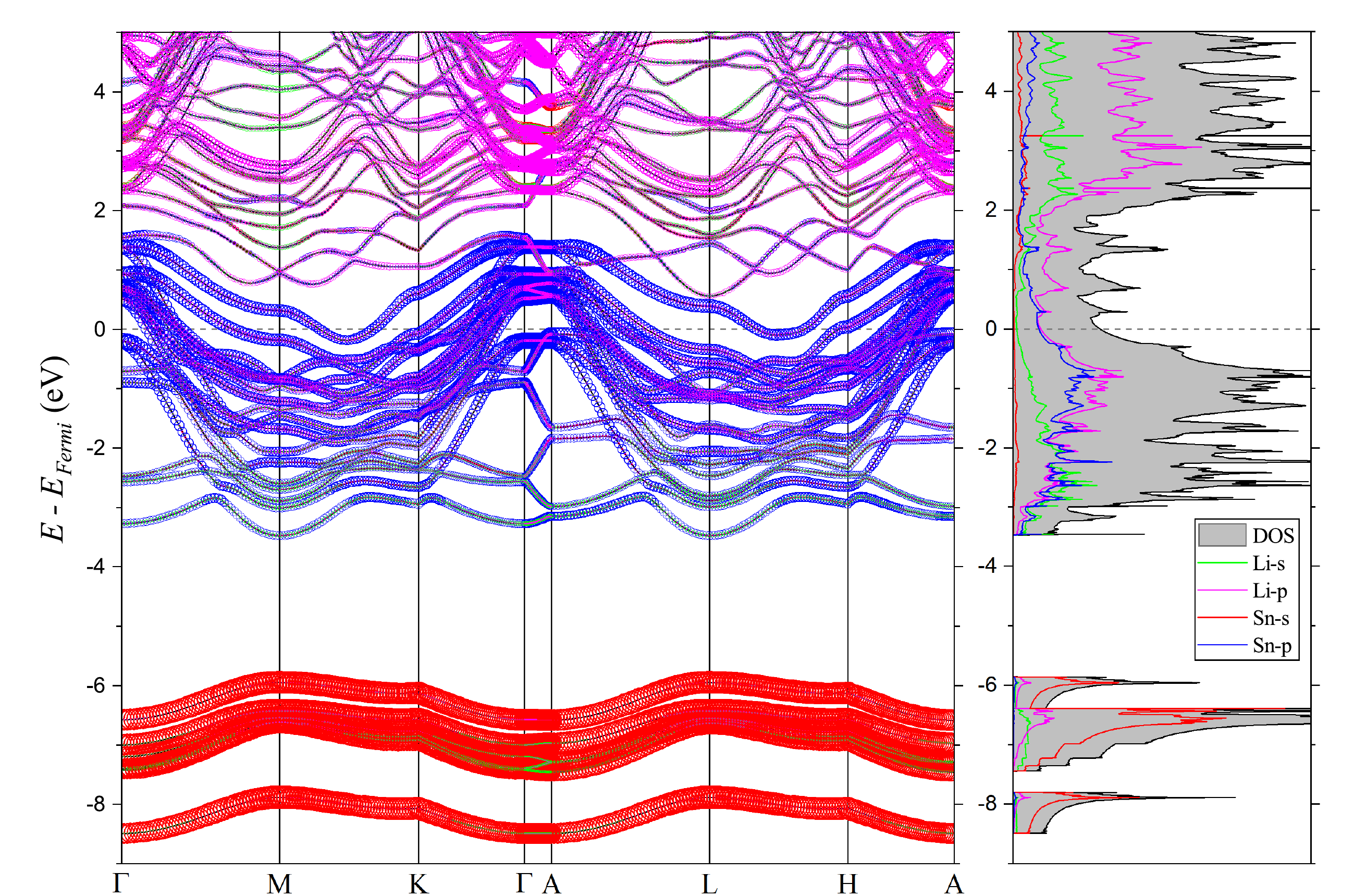}
  \caption{Band structure and density of states (DOS) for hR75-Li$_{19}$Sn$_6$ at $P=0$ GPa.}
  \label{figS6}
\end{figure}

\begin{figure}[ht]
\centering
  \includegraphics[width=0.8\textwidth]{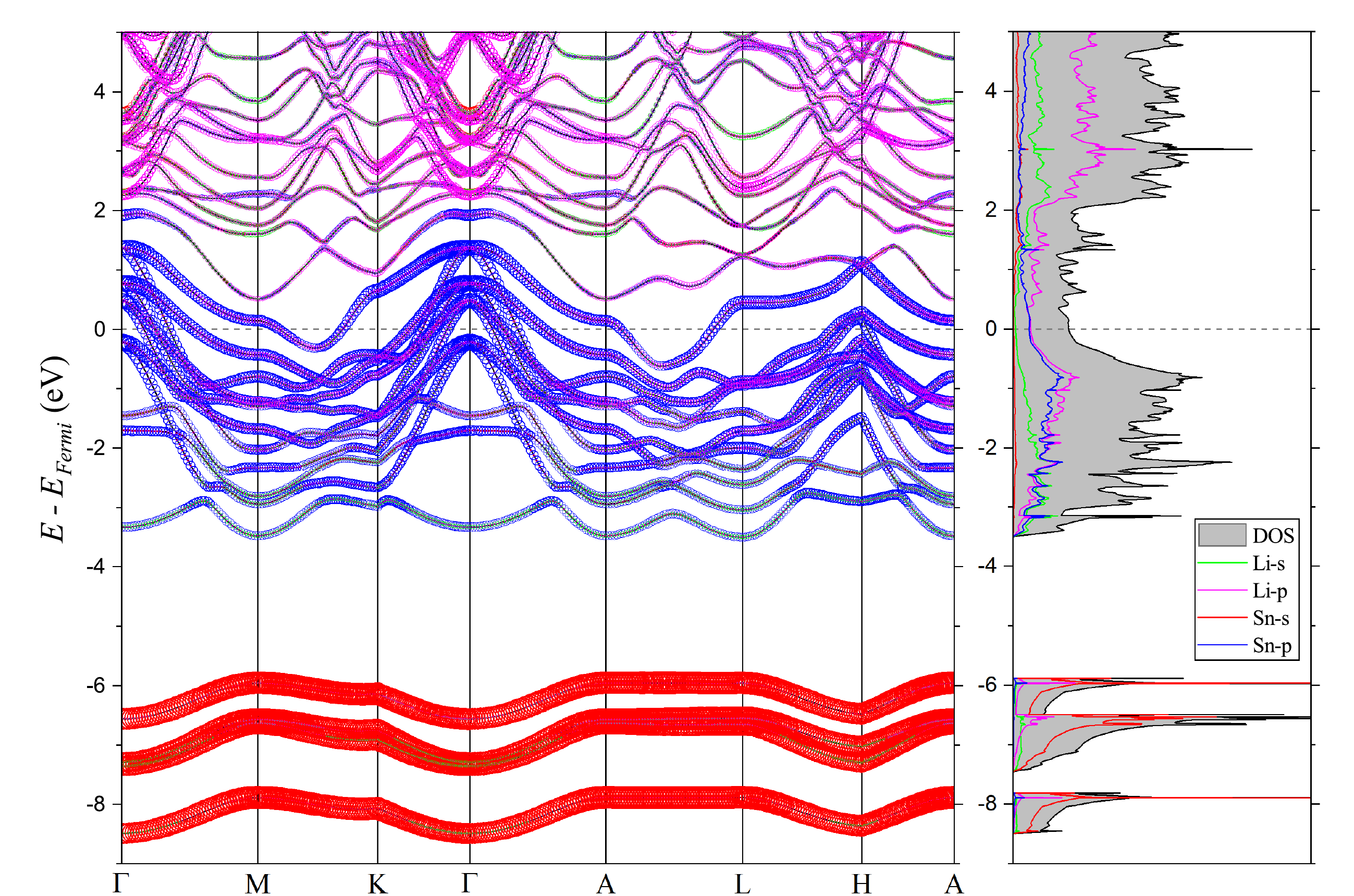}
  \caption{Band structure and density of states (DOS) for hR48-Li$_3$Sn at $P=0$ GPa.}
  \label{figS7}
\end{figure}

\begin{figure}[ht]
\centering
  \includegraphics[width=0.9\textwidth]{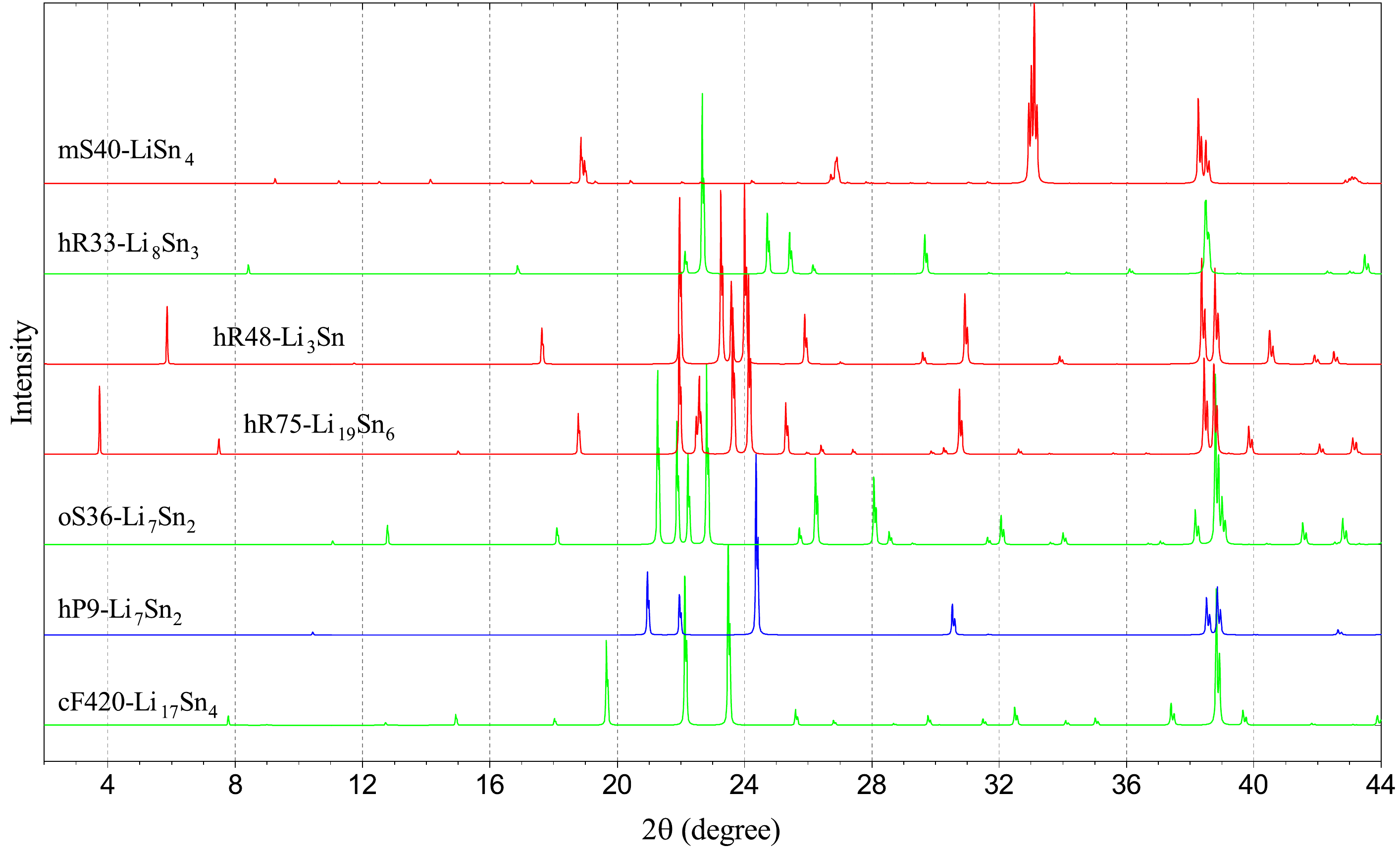}
  \caption{ Powder x-ray diffraction patterns simulated for 
  $\lambda_{\text{Cu K}-\alpha}= 1.54059$ \AA\ for some of the known and predicted
  ambient-pressure Li-Sn phases.
  } 
\label{figS2} 
\end{figure}
